\documentclass[journal]{IEEEtran}
\usepackage{xcolor,cite,etoolbox}
\usepackage{balance}
\makeatletter 
\usepackage{amsmath,amssymb,amsfonts}
\usepackage{algorithmic}
\usepackage{algorithm}
\usepackage{graphicx}
\usepackage{color}
\usepackage{bm}
\usepackage{subfigure}
\usepackage{diagbox}

\usepackage{url}

\begin{document}

\title{Edge Federation: Towards an Integrated Service Provisioning Model}

\author{Xiaofeng~Cao,~\IEEEmembership{Student Member,~IEEE,}
		Guoming~Tang,~\IEEEmembership{Member,~IEEE,} 
        Deke~Guo,~\IEEEmembership{Member,~IEEE,~ACM,} 
        Yan~Li, 
        and~Weiming~Zhang % <-this % stops a space
\thanks{X. Cao, D. Guo, Y. Li, W. Zhang are with the Science and Technology on Information Systems Engineering Laboratory, National University of Defense Technology, Changsha, Hunan, 410073, P. R. China (e-mail: \{caoxiaofeng10, dekeguo, liyan10, wmzhang\}@nudt.edu.cn). G. Tang is with the Peng Cheng Laboratory, Shenzhen, Guangdong 518055, P. R. China. (e-mail: tanggm@pcl.ac.cn).
\newline Corresponding author: Deke Guo. }% <-this % stops a space
}

\maketitle
\begin{abstract}
Edge computing is a promising computing paradigm by pushing the cloud service to the network edge. To this end, edge infrastructure providers (EIPs) need to bring computation and storage resources to the network edge and allow edge service providers (ESPs) to provision latency-critical services for end users. Currently, EIPs prefer to establish a series of private edge-computing environments to serve specific requirements of users. This kind of resource provisioning mechanism severely limits the development and spread of edge computing in serving diverse user requirements. In this paper, we propose an integrated resource provisioning model, named \emph{edge federation}, to seamlessly realize the resource cooperation and service provisioning across standalone edge computing providers and clouds. To efficiently schedule and utilize the resources across multiple EIPs, we systematically characterize the provisioning process as a large-scale linear programming (LP) problem and transform it into an easily solved form. Accordingly, we design a dynamic algorithm to tackle the varying service demands from users. We conduct extensive experiments over the base station networks in Toronto. Compared with the fixed contract model and multihoming model, edge federation can reduce the overall costs of EIPs by 23.3\% to 24.5\%, and 15.5\% to 16.3\%, respectively.
\end{abstract}


\section{Introduction}\label{sec:intro}

The emergence of edge computing offers a new paradigm to deliver computation-intensive and latency-critical services to traditional cloud users~\cite{gsma,Guoming}. The basic idea is to push the cloud service to the network edge (e.g., access points, base stations, or central offices), which is closer to users than the cloud. In this way, users can still exploit the power of cloud computing, while no longer suffering from network congestion and long latency~\cite{shi2016edge}. The prosperous development of edge computing offers an excellent opportunity for service providers, where they can rent resources from edge infrastructure providers (EIPs) to host their services. An EIP usually has to deploy and maintain a set of distributed edge nodes at the network edge, where an edge node can consist of multiple edge servers and be configured with certain computation and storage resources. As the owners of edge nodes, EIPs are responsible for service provisioning and resource management.

Nevertheless, compared with the existing cloud computing solution, edge computing is still constrained in the resource capacity and implemented with high costs in maintaining edge infrastructure widely~\cite{shi2016edge},~\cite{sdxedge}. The main reason is that EIPs tend to establish a series of private edge-computing environments to serve specific requirements of users from the aspect of their own viewpoint~\cite{wang2017jointcloud}. That is, each EIP only manages and uses its own resources; hence, a standalone edge-computing environment is usually resource-constrained, especially in the scenario of serving the increasing amount of users. When a large number of services need to be deployed in broad geographic areas, the involved EIPs have to deploy and maintain more edge nodes for service coverage potentially leading to a huge cost. On the other hand, however, different EIPs may build edge nodes in the same place independently without any cooperation, causing unbalanced and under-utilized edge resources. To make matters worse, since the individual EIP has limited information about the whole edge-computing environment, it is tough to make a global optimization for efficiently delivering various services to different users. Consequently, the inefficient resource management and service deployment paradigm could severely limit the development of a healthy edge computing ecosystem.

With the above challenges in mind, this paper presents \emph{edge federation}, an integrated service provisioning model for the edge computing paradigm. It aims to establish a cost-efficient platform for EIPs and offer end users and ESPs a transparent resource management scheme by seamlessly integrating individual EIPs as well as clouds.

\textbf{Private or Public:} In the horizontal dimension, EIPs independently construct and maintain their private resource infrastructures, which restricts the quick development and spread of edge computing. In the existing model, an EIP usually has a limited amount of edge servers, and hence cannot cover broad areas and may cause long service delivery latency to those users outside the covered areas. This would severely limit the market size of each EIP. A straightforward solution for the individual EIP is to build edge servers at more locations. This method, however, would cause a large amount of duplicated edge nodes across EIPs in many sites, leading to the increased capital and operational expenditure. Therefore, it is sensible to enable interoperability and resource sharing across EIPs.

\textbf{Edge or Cloud:} In the vertical dimension, cloud computing and edge computing both have their own advantages, but neither of them can meet the high latency requirement (a.k.a., low time delay) of services and low-cost resource provision simultaneously. Although edge computing can achieve much lower latency in service delivery than cloud computing, it also incurs a high deployment cost of new computation and storage infrastructures~\cite{Amazon},~\cite{Google}. On the contrary, the low cost and sufficient resources are precisely the advantages of cloud computing. Moreover, as each edge node has a limited range of serving areas, the cloud could be an essential complement to support end users in the areas not served by edge nodes~\cite{ma2017cost}. In summary, edge computing and cloud computing can reasonably complement each other, with an effective mechanism to cooperate.

The edge federation proposed in this paper brings a new service-provisioning model for the next generation edge-computing network, which could be a triple-win solution for EIPs, ESPs and end users. For EIPs, more effective service deployment and delivery can be achieved by fewer infrastructure constructions, resulting in higher revenue. For ESPs, due to the shared resource pool in edge federation, the market size could be easier to expand, and the reliability of service delivery can be considerably enhanced. To this end, end users can enjoy an improved service experience with lower latency.
To realize edge federation, we are facing three critical challenges.
First, an edge-computing network is very complex, which consists of a series of EIPs, diverse services, and heterogeneous end devices. The edge federation designed in this paper should realize the targets of scalability, efficiency, and low-latency.
Second, the edge federation should effectively model the joint service provisioning process across heterogeneous edge nodes, and across edge and cloud.
Third, the service provisioning problem under the edge federation involves a large number of optimization variables and exhibits very high computation complexity. An efficient solution with affordable complexity is needed to deal with the large-scale optimization problem.

We address all the above challenges in this work and make the following major contributions:

\begin{itemize}

\item{We design the edge federation, an integrated edge computing model, to realize the transparent service provisioning across independent EIPs and the cloud. The edge federation model is designed to improve the QoE of end users and save the cost of EIPs.}

\item{We characterize the service provisioning process under our edge federation as a linear programming (LP) optimization model and propose a dimension-shrinking method to reformulate it into an easily solved model. Accordingly, we develop a service provisioning algorithm \emph{SEE}.}

\item{We evaluate the proposed solution for edge federation under the base station network of Toronto city with the real-world trace. Compared with the fixed contract model and multihoming model, edge federation can reduce the overall cost of EIPs by 23.3\% to 24.5\%, and 15.5\% to 16.3\%, respectively.}

\end{itemize}

The rest of the paper is organized as follows. We introduce the background and the related challenges of the edge federation in Sec.~\ref{sec:background}. Then, the detailed architecture is illustrated, and the contributions are also highlighted in Sec.~\ref{sec:framework}. Sec.~\ref{Sec:formulation} formulates the cost minimization problem for EIPs with hard latency constraints. In Sec.~\ref{sec:algorithm}, the problem is transformed with the dimension-shrinking method and the dynamic provisioning algorithm is developed. We evaluate the performance of our solution using real-world network service data and validate the effectiveness of the edge federation model in Sec.~\ref{sec:eva}. Sec.~\ref{sec:discussion} gives the discussion and future work. Sec.~\ref{related work} reviews the related work and the state-of-the-art. Sec.~\ref{conclusion} concludes the paper.

\section{Edge Federation vs. Cloud Federation}\label{sec:background}

Edge federation is the platform that spans the continuum of resources in different EIPs, from cloud to the edge. In a cross-EIP way, edge federation can bring the customized resources (e.g., computation, storage, and networking resources) for ESPs and end users in a broad, fast, and reliable geo-distributed manner. 

A similar idea to the edge federation is the cross-cloud cooperation architecture in previous works. Such works attempted to establish the integrated cloud resources provisioning architecture, named as Joint Cloud~\cite{wang2017jointcloud}, Hybrid Cloud~\cite{hybridcloud}, etc. The cloud federation tries to establish the environment that combines the public and the private resources, which can enable infrastructure providers scale resources for handling short-term spikes (e.g., Black Friday in the Amazon, Single's day in the Taobao, etc.) in demand~\cite{wang2017jointcloud}. It can be regarded as the horizontal integration mentioned before. Some works also considered vertical integration in the field of content caching or computation offloading. Most of the works constructed the cloud-assisted~\cite{ma2017cost} or edge-assisted~\cite{yang2018cost} network structures, both of which aimed to solve two main problems: the limitation of the edge resource capacity and the long latency caused by the backhaul network from users to the cloud.

Compared with the aforementioned works, edge federation is much different and the construction is even more challenging. The resource integration in the edge federation could be more complicated and urgent, mainly due to three aspects of the edge computing: (i) the highly distributed, limited and heterogeneous edge resources, (ii) the high cost of edge resources, and (iii) the latency-critical and computation-intensive edge services. Based on these characteristics, we are facing several particular challenges in edge federation.

\begin{enumerate}

\item{\emph{The trade-off between the cloud and the edge:} As described in the previous section, the edge can achieve the lower service latency but with higher cost. In contrast, the cloud may incur a lower cost but with higher latency. Neither of them can meet the high latency requirement of services and low-cost resource provision simultaneously. Thus, the goal of the edge federation is trying to strike a balance between the latency and the cost, either the trade-off between the cloud and the edge. How to use the least cost to fulfill the service requirements and achieve the best QoS is the most critical problem in the edge federation.}

\item{\emph{The optimization of resource allocation on distributed and limited edge resources:} Compared with cloud nodes, edge nodes are much more scattered in geography with a limited amount of resources. Due to such the limitation, EIPs have to be careful when they provide the resource to services. This severely restricts their capacity in the size of the serving area and service demands. Thus, the cooperation of different EIPs and the optimization of resource provision in the edge federation are more urgent than those in the cloud computing scenario. The challenge is how to maximize the resource provisioning efficiency in the highly distributed and limited edge resources.}

\item{\emph{The contradiction between the computation-intensive edge services and the limited edge resources:} The resources in edge nodes are limited. Worse still, most of the emerging services in the edge scenario have high computation and strict latency requirements (e.g., self-driving services, virtual reality, etc.) which require significant computation and storage resources. This dilemma makes the edge more likely to get into the ``Spike" trouble (i.e., overload trouble) and suffer from resource shortages.}

\end{enumerate}

For these challenges, an efficient service provisioning method is needed. In the following section, we first design the architecture of the edge federation, under which the corresponding service provisioning method can be developed.

\section{THE ARCHITECTURE OF EDGE FEDERATION}\label{sec:framework}
We start with an initial example and an overview of the edge federation and then present the detailed architecture of the edge federation.

\subsection{Rationale}

\begin{figure}
\centerline{\includegraphics[width=9cm]{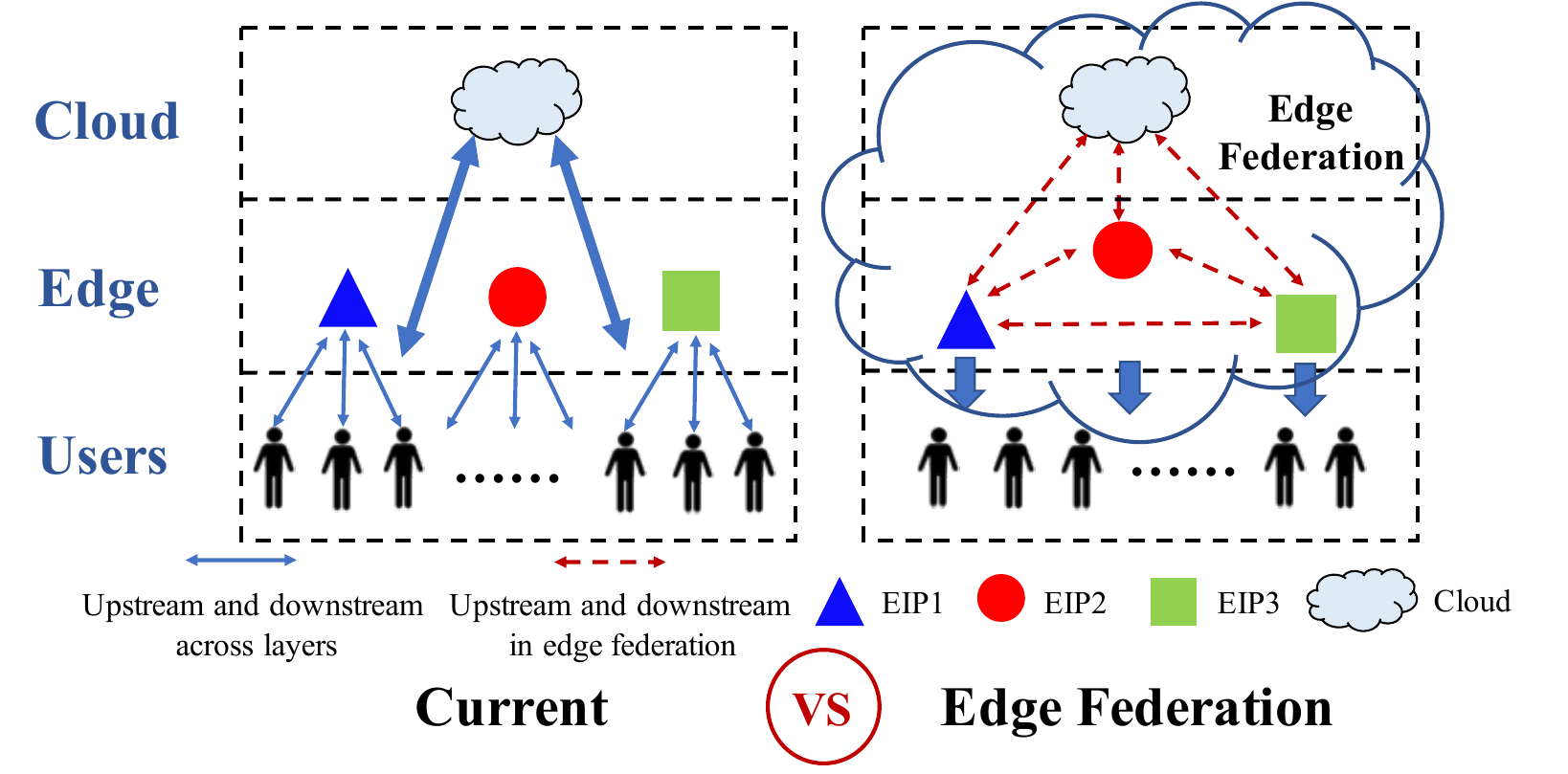}}
\vspace{-0.1in}
\caption{The comparison between the current edge computing paradigm and the edge federation. The edge federation consortium can realize a transparent resource management mechanism over EIPs and cloud.}
\label{fig:comparsion future and current}
\vspace{-0.2in}
\end{figure}

As shown in the left of Fig.~\ref{fig:comparsion future and current}, existing network environment mainly has three layers: (i) the user layer consists of a large amount of heterogeneous smart devices (e.g., mobile phones, vehicles, etc.), which dynamically request for high-performance services from ESPs; (ii) the edge layer is formed by EIPs, which are responsible for resource provisioning for ESPs. EIPs provide computation and storage resources, as well as techniques (e.g., NFV and SDN) and platforms (e.g., Amazon Web Services, Microsoft Azure ,etc.); (iii) the cloud layer provides similar types but a larger amount of resources to end users. In the current network environment, ESPs usually sign contracts and package the content of their services to EIPs. An EIP usually manages its own resources and deliver the contracted services to corresponding end users.

Current service-provisioning models for individual EIP can be inefficient and costly. \textbf{From the perspective of resources}, EIPs independently deploy edge nodes at the edge of the network, where each edge node consisting of multiple edge servers provides computation and storage resources for accommodating diverse services. The capacity and the serving range of an individual edge node are much smaller than those of the cloud. Moreover, EIPs independently manage their resources without any cooperation in the current edge-computing model. Consequently, such a mechanism fails to achieve globally optimal scheduling of resources and services, hence leading to the resource overloaded or under-utilization situation and even resulting in a low QoS. \textbf{From the perspective of cost}, each EIP tends to build more edge nodes in new locations to increase the amount of resources and expand the service coverage. Multiple EIPs even build edge nodes in the same location for the market competition. Such a method would cause a huge overhead (e.g., the expenditure of infrastructure construction, maintenance cost, energy cost, etc.) and severe resource waste. Eventually, such heavy burdens will be taken by either EIPs, ESPs, or end users in this triple-lose situation.

To overcome the above disadvantages, we propose edge federation, a transparent service-provisioning model in the multi-EIP environment. It involves two-dimension integration for the service deployment, including the integration between edge and cloud, and the seamless cooperation among heterogeneous edge nodes of multiple EIPs. The basic idea of edge federation is shown on the right side of Fig.~\ref{fig:comparsion future and current}, where each EIP and cloud is a member of edge federation, and the edge nodes and cloud nodes can share resources and interact with each other, where they are not necessary to be genuinely interconnected. They only disclose detailed information to the authoritative and trusted consortium, which is the core of edge federation.

\subsection{Architecture of Edge Federation}

\begin{figure}
\centerline{\includegraphics[width=8cm]{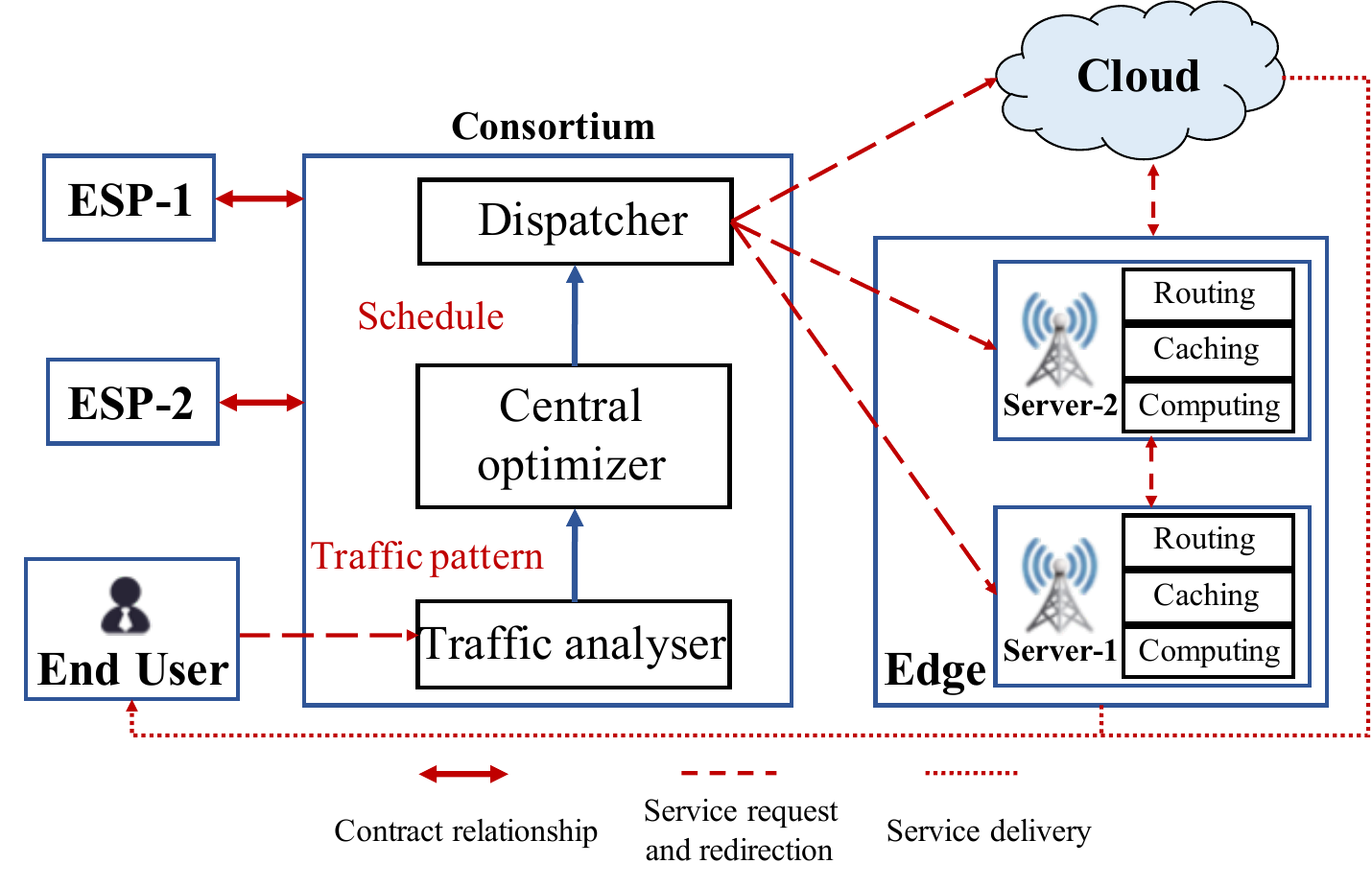}}
\caption{The architecture of edge federation.}
\label{fig:architecture-of-EF}
\vspace{-0.2in}
\end{figure}

As shown in Fig.~\ref{fig:architecture-of-EF}, edge federation consortium mainly consists of three components, including the traffic analyzer, the central optimizer, and the dispatcher.

\emph{Traffic analyzer} is a module that continuously analyzes the traffic pattern based on various service requests from end users. The traffic patterns can accurately characterize the service demands temporally and spatially and will be served as an essential input to the central optimizer. Considering that many proposals have devoted to traffic prediction and modeling, we use the existing methods\footnote{The short-term prediction (e.g., conventional methods: ARIMA~\cite{calheiros2015workload}, etc., computation intelligence methods with off-line training and on-line prediction: LSTM~\cite{laptev2017time} etc..) has been intensely developed and proven to be reliable with high prediction accuracy. They can provide proper input to the further optimization.} (e.g., ARIMA~\cite{calheiros2015workload}) in our traffic analyzer to predict the traffic. A comprehensive study of the traffic prediction and modeling is out of the scope of this paper.

\emph{Central optimizer} is the brain of edge federation consortium. It computes the traffic redirection schedule based on the obtained traffic pattern and the temporal-spatio information of end users (e.g., location, time, type of requesting service). Based on the schedule, EIPs deploy the corresponding requested services at the edge and cloud, and the dispatcher will redirect the service request accordingly.

\begin{figure}[t]
\centering
\includegraphics[width=8cm]{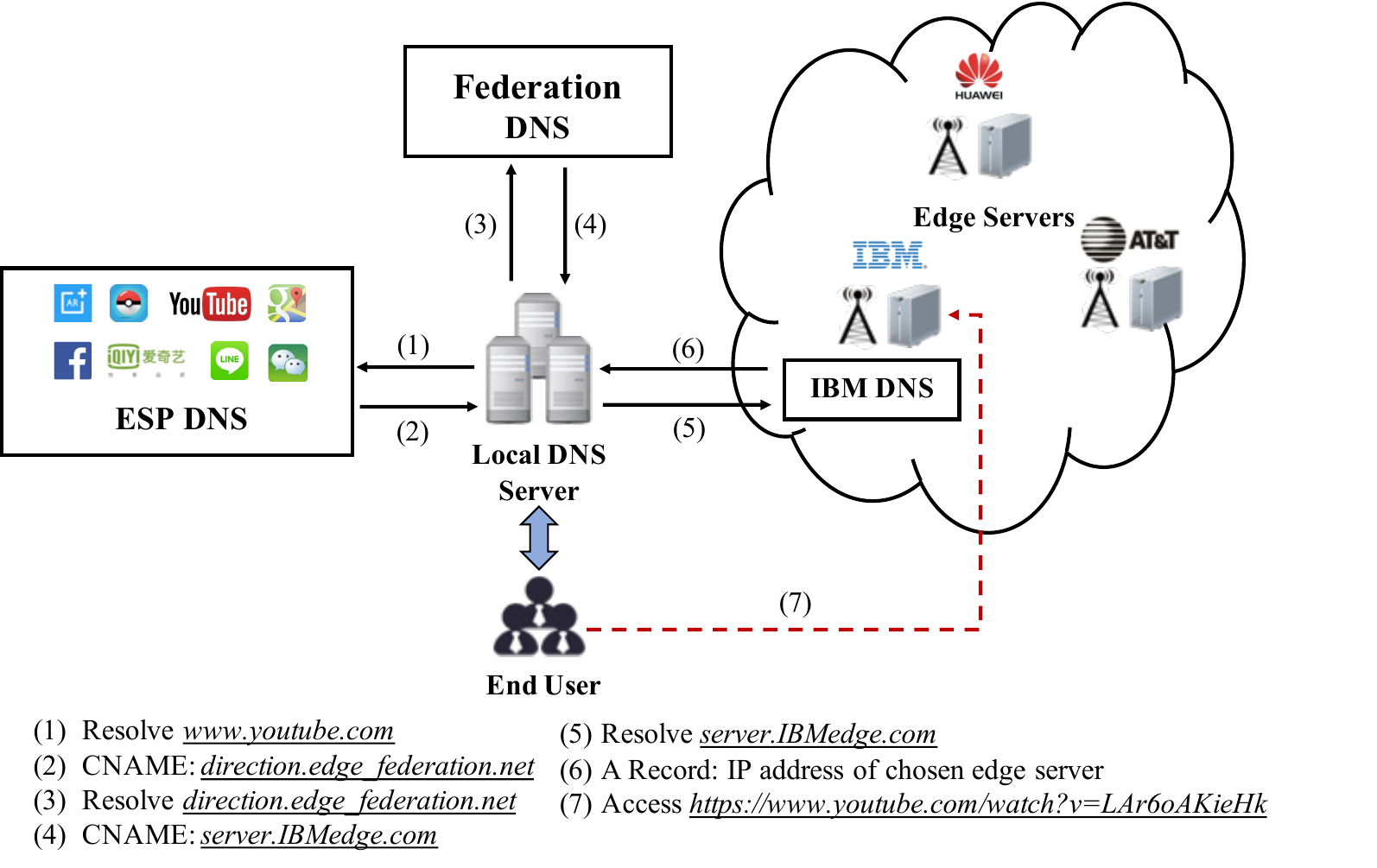}
\caption{An example of request direction.}
\label{fig:dns}
\vspace{-0.2in}
\end{figure}

\emph{Dispatcher} redirects users' service requests to correct edge or cloud servers. Such redirection can be performed by the existing routing mechanisms (DNS CNAME record, A record). To ease the understanding, we present a detailed example of service redirection based on the DNS technique in Fig.~\ref{fig:dns}. The end user at a specific area requests a video of \emph{YouTube}. Different from the traditional mechanism, the EIP DNS modifies its CNAME record to point to the domain of a federation DNS instead of the contracted EIP DNS domain. Then, based on the CNAME record from the redirection schedule, the consortium dispatcher makes the decision to redirect the user's request to the optimal edge server. Thus the high-performance service can be achieved. 

\subsection{Benefits of Edge Federation}
\subsubsection*{The business relationship} Traditionally, ESPs will run services on the infrastructure of EIPs with the pay-as-you-go function. Different EIPs will manage their resources and deliver services to the end user independently. The difference between the money ESPs paid to EIPs, and the operation cost (e.g., storage cost, computation cost, communication cost, etc.) of EIPs is the revenue of the EIP. In the edge federation, ESPs will also run services on the EIP and pay the corresponding fee to the EIP. However, these services will be deployed by the edge federation with a global view of the unified resource pool, which consists of cloud nodes and edge nodes from different EIPs. Then, the node will deliver the corresponding service to the end user.

\subsubsection*{1) For EIPs} EIPs in the conventional model can only manage the corresponding service delivery on their edge nodes in limited areas. Compared with the traditional method, edge federation makes it possible that EIPs can operate the service more flexible among the unified resource pool. Such the method can help EIPs deliver the service to end users with a shorter distance, less infrastructure construction, and thus enable a more cost-efficient service deployment and delivery with reasonable edge cooperation and cloud assist. Therefore, the operation cost of the EIP can be reduced, and the revenue of the EIP can be improved.

\subsubsection*{2) For ESPs} In the existing method, due to the limited coverage area of a single EIP, the corresponding contracted ESP can only spread its service in a considerably small region, which means that the ESP has a limited market size. Fortunately, due to the wide and densely distributed edge nodes of different EIPs in the unified resource pool, the above dilemma will not exist in the edge federation. Moreover, with the same unit price, ESPs will get a higher QoS.

\subsubsection*{3) For end users} Edge federation makes ESPs run their services on any edge nodes of multiple EIPs. These edge nodes can be distributed in a variety of geographical locations. As a result, end users can potentially get the services from closer nodes with lower latency. Moreover, the reliability of service delivery can also be considerably enhanced.

Note that edge federation is a trusted consortium, which is authorized to control the resources of EIPs. For the existing infrastructures, edge federation mainly aims to union the infrastructures of different EIPs to improve service performance via optimized resource management and service deployment. For further business development and extension, edge federation could be a joint-stock company, which shall undertake tasks of not only global resource management but also the infrastructure construction. Thus, the infrastructure can be the shared facility instead of belonging to a certain EIP and is only managed by the edge federation. For the detailed design of cooperation paradigm or models cross different EIPs, however, it falls into the field of Network Economics and is beyond the scope of this paper.
As to the privacy issue raised by user data sharing among EIPs, previous work can be found in tackling it, e.g., private information retrieve~\cite{sun2017capacity} and data encryption~\cite{gai2017privacy}. In addition, previous works also demonstrated that data privacy could be guaranteed under the cloud computing and big data scenario, even the infrastructure provider is unrelable~\cite{naveed2014dynamic}.

\section{Optimal Service Provisioning in Edge Federation}\label{Sec:formulation}

After introducing the architecture of edge federation, in this section, we present the service provisioning process in detail. We first model the dynamic service demands, based on which the two-phase resource allocation process is formulated from the vertical and horizontal dimensions. Then, in order to guarantee the service performance, we get the latency constraints and formulate the cost minimization problem of edge federation.

\newcommand{\tabincell}[2]{\begin{tabular}{@{}#1@{}}#2\end{tabular}}
\begin{table*}[ht]
\centering
\caption{Main notations}
\vspace{-0.12in}
\label{my-label}
\centering
\begin{tabular}{|l|l|}
\hline
Notation & Description\\ \hline
\hline
$T$    & A time period of $n$ consecutive time slots.                           \\ \hline
$U$    & Set of end users.               \\ \hline
$P$    & Set of ESPs.                \\ \hline
$A$    & Set of cloud nodes $a$. \\ \hline
$E$    & Set of edge nodes $e$ in EIPs.\\ \hline
\hline
$\alpha_{u,p}^{e}(t)$ & \tabincell{l}{Fraction of storage demands of service $p$ from end user $u$ assigned to edge node $e$ at time slot $t$.} \\ \hline
$\beta_{u,p}^{e}(t)$ & \tabincell{l}{Fraction of computation demands of service $p$ from end user $u$ assigned to edge node $e$ at time slot $t$.} \\ \hline
$\theta_{u,p}^{S,a}(t)$ &{Fraction of storage demands of service $p$ from end user $u$ assigned to cloud $a$ at time slot $t$.} \\ \hline
$\theta_{u,p}^{C,a}(t)$ &{Fraction of computation demands of service $p$ from end user $u$ assigned to cloud $a$ at time slot $t$.} \\ \hline

$S_{u,p}(t)$  & \tabincell{l}{Amount of storage demands of service $p$ from end user $u$ at time slot $t$ before computation.} \\ \hline
$S_{u,p}^{'}(t)$  & \tabincell{l}{Amount of delivery contents of service $p$ from end user $u$ at time $t$ after computation.} \\ \hline
$C_{u,p}(t)$ & Amount of computation demands of service $p$ from end user $u$ at time slot $t$. \\ \hline
$S_{a}$ & \tabincell{l}{Storage capacity of cloud node $a$.}                                         \\ \hline
$C_{a}$ & \tabincell{l}{Computation capacity of cloud node $a$.}                              \\ \hline
$S_{e}$ & \tabincell{l}{Storage capacity of edge node $e$.}                                         \\ \hline
$C_{e}$ & \tabincell{l}{Computation capacity of edge node $e$.}                              \\ \hline
$S_{E}(t)$ & Total amount of storage demands deployed on edge at time slot $t$.\\ \hline
$C_{E}(t)$ & Total amount of computation demands deployed on edge at time slot $t$.\\ \hline
\hline
$l_{u,p}(t)$ & Service latency of service $p$ for end user $u$ at time slot $t$. \\ \hline
$h_{u}^{a}$ & Delivery distance between cloud server $a$ and end user $u$. \\ \hline
$h_{u}^{e}$ & Delivery distance between edge server $e$ and end user $u$. \\ \hline
$l_p$ & Latency requirement of service $p$. \\ \hline
$m_{u,p}(t)$ & \tabincell{l}{Service satisfaction parameter indicates whether the service meets the latency requirement.} \\ \hline
$l_{p}^{sat}$ & Satisfaction ratio of specific service $p$. \\
\hline
\end{tabular}
\vspace{-0.2in}
\end{table*}

\subsection{Modeling the Service Provisioning}

\emph{1) Network Environment and Dynamic Service Demands}

For an edge-computing network, there are various edge nodes, each of which may consist of multiple edge servers. We assume that end users are geographically distributed according to the locations of edge nodes. Generally, there are four roles in the entire edge-computing network. Define $U$ as the set of all end users, $A$ as the set of cloud nodes, $E$ as the set of edge nodes, and $P$ as the set of edge services, respectively. Let $u$ $\in$ $U$ represents a specific user, $a$ $\in$ $A$ represents a specific cloud node, $e$ $\in$ $E$ represents a specific edge node, while $p$ $\in$ $P$ represents a specific edge service. For simplicity, we assume that the topology of the designed edge federation is known in advance. The main notations are shown in Table.~\ref{my-label}.

The end users have time-varying service demands toward the storage and computations. The service demands within a time period $T$ could be divided into $n$ smaller equal time slots, e.g., 1 hour. Let the service demands $p$ from end user $u$ at time slot $t$ be $K_{u,p}(t)={\{S_{u,p}(t),S_{u,p}^{'}(t),C_{u,p}(t)}\}$ ($\forall t\in T, t = {1,2,\dots,n}$). Here, $S_{u,p}(t)$ and $S_{u,p}^{'}(t)$ represent the amount of content before and after processing, respectively, while $C_{u,p}(t)$ denotes the computation demands of accomplishing the service. These terms can be captured as follows:

\begin{equation}
    \sum\limits_{u \in U}S_{u,p}(t) = |U| \cdot q_p(t), \forall p \in P,
    \label{generate 1}
\end{equation}
\begin{equation}
    S_{u,p}^{'}(t) = S_{u,p}(t) \cdot k_s, \forall u \in U, \forall p \in P, \forall t \in T,
    \label{generate 2}
\end{equation}
\begin{equation}
    C_{u,p}(t) =  S_{u,p}(t) \cdot k_c, \forall u \in U, \forall p \in P, \forall t \in T,
    \label{generate 3}
\end{equation}
where $|U|$ refers to the population of the target location area, $q_p(t)$ is the normalized traffic profile at time slot $t$ whose value is related to a specified service $p$. $k_s$ is the coefficient profile that describes the size of the content after processing, and  $k_c$ is the coefficient profile that describes the required computation resource for accomplishing the corresponding task. The traffic demands of the service $p$ in the related area around edge server $e$ at time $t$ can be also captured by:
\begin{equation}
    d_{ep}(t)=|U|_e \cdot q_p(t), \forall p \in P,
\end{equation}
where $|U|_e$ represents the population of specified edge server location. As the service demands of users are highly dynamic, a tailored time slot is necessary to the cost-efficiency schedule of the edge federation. Our implementation in later experiments shows that the time slot with a length of 1 hour is good enough to yield the better result than the existing method. We also give a discussion on how to choose the suitable length of the time slot in Sec. VII.

\emph{2) Two-phase Resource Allocation}

Vertically, we assume that each EIP will resolves part or all the storage demands $S_{u,p}(t)$ and computation demands $C_{u.p}(t)$ by cloud nodes. Two variables $\theta_{u,p}^{S,a}(t)$ and $\theta_{u,p}^{C,a}(t)$ represent the fraction of storage and computation demands supplied by cloud node $a$ at time slot $t$, respectively. The other $(1-\sum\limits_{a \in A}\theta_{u,p}^{S,a}(t))$ storage and $(1-\sum\limits_{a \in A}\theta_{u,p}^{C,a}(t))$ computation demands will be served by the edge nodes. The values of these fractions fall in the range of $[0,1]$:
\begin{equation}
    0\leq \theta^{S,a}_{u,p}(t)\leq 1 ,  \forall u \in U, \forall p \in P, \forall t \in T,
\label{constraint1}
\end{equation}
\begin{equation}
    0\leq \theta^{C,a}_{u,p}(t)\leq 1 ,  \forall u \in U, \forall p \in P, \forall t \in T.
\label{constraint2}
\end{equation}

At any time slot $t$, the storage and computation demand should not exceed the capacity of the involved cloud nodes and edge nodes. Thus, we have the following two constraints:
\begin{equation}
    \sum\limits_{u \in U}\sum\limits_{p \in P}S_{u,p}(t)\theta_{u,p}^{S,a}(t)\leq S_a , \forall t \in T,
\label{constraint3}
\end{equation}
\begin{equation}
    \sum\limits_{u \in U}\sum\limits_{p \in P}C_{u,p}(t)\theta_{u,p}^{C,a}(t)\leq C_a , \forall t \in T,
\label{constraint4}
\end{equation}
where $S_a$ and $C_a$ are the storage and computation capacity of a specific cloud node $a$. Given the dynamic demands from each location area of end users, we assume that the least cloud capacities should be able to satisfy the peak service demands, thus:
\begin{equation}
    \sum\limits_{a \in A}S_a= \max\limits_{t \in T}\{\sum\limits_{u \in U, p \in P, a \in A}S_{u,p}(t)\theta_{u,p}^{S,a}(t)\},
\end{equation}
\begin{equation}
    \sum\limits_{a \in A}C_a= \max\limits_{t \in T}\{\sum\limits_{u \in U, p \in P, a \in A}C_{u,p}(t)\theta_{u,p}^{C,a}(t)\}.
\end{equation}

Noting that the service demands of end users may change over time, so the edge federation needs to manage available resources hourly. 
Compared to accessing an edge node, the latency of accessing a cloud node is higher, however, with lower resource cost. Therefore, to achieve a trade-off between the edge and cloud (i.e., the cost and latency), we need to dynamically find the reasonable variables $\theta_{u,p}^{S,a}(t)$ and $\theta_{u,p}^{C,a}(t)$ by the management of the edge federation.

Horizontally, the amount of storage and computation demands supplied by edge nodes are:
\begin{equation}
    S_E(t)= \sum\limits_{u \in U}\sum\limits_{p \in P}S_{u,p}(t)(1-\sum\limits_{a \in A}\theta_{u,p}^{S,a}(t)),
\end{equation}
\begin{equation}
    C_E(t)= \sum\limits_{u \in U}\sum\limits_{p \in P}C_{u,p}(t)(1-\sum\limits_{a \in A}\theta_{u,p}^{C,a}(t)).
\end{equation}
 
We use two variables $\alpha_{u,p}^{e}(t)$ and $\beta_{u,p}^{e}(t)$ to denote the fraction of storage demands $S_{u,p}(t)$ and computation demands $C_{u,p}(t)$ supplied by edge node $e$ at time slot $t$, respectively. Thus, we have the following constraints:
\begin{equation}
    0\leq \alpha_{u,p}^{e}(t)\leq 1 ,  \forall u \in U,\forall p \in P,\forall e \in E,\forall t \in T,
\label{constraint5}
\end{equation}
\begin{equation}
    0\leq \beta_{u,p}^{e}(t)\leq 1 ,  \forall u \in U,\forall p \in P,\forall e \in E,\forall t \in T.
\end{equation}

Then, we define the storage and computation capacity of the edge node $e$ as $S_e$ and $C_e$, respectively. They represent the maximum demands the edge node can serve in a single time slot. Then, we have the following constraints:
\begin{equation}
    \sum\limits_{u \in U}\sum\limits_{p \in P}S_{u,p}(t)\alpha_{u,p}^{e}(t)\leq S_e , \forall e \in E,\forall t \in T,
\label{formula1}
\end{equation}

\begin{equation}
    \sum\limits_{u \in U}\sum\limits_{p \in P}C_{u,p}(t)\beta_{u,p}^{e}(t)\leq C_e , \forall e \in E,\forall t \in T.
\label{formula2}
\end{equation}

Formulas (\ref{formula1}) and (\ref{formula2}) mean that the storage and computation demands assigned to the edge node $e$ should not exceed its capacity at any time slot. Ideally, those demands from all users should be completely satisfied. Thus, we have
\begin{equation}
    \sum\limits_{e \in E}\alpha_{u,p}^{e}(t)+\sum\limits_{a \in A}\theta_{u,p}^{S,a}(t)=1 , \forall{u \in U}, \forall p \in P, \forall t \in T,
\end{equation}
\begin{equation}
    \sum\limits_{e \in E}\beta_{u,p}^{e}(t)+\sum\limits_{u \in U}\theta_{u,p}^{C,a}(t)=1 , \forall{u \in U},  \forall p \in P, \forall t \in T.
\label{constraint6}
\end{equation}

\emph{3) Cost Minimization for the Edge Federation}

To serve the demands of a set of users, the edge federation treats the minimization of the overall cost (i.e., maximization of revenue) of EIPs as an important optimization goal. Under the edge-computing scenario, the overall cost, $V$, can be divided into three parts, including the computation cost, the storage cost, and the communication cost. 

\subsubsection*{\textbf{Remark 1}} {\color{black}\emph{For the cost of EIPs, it could be roughly divided into two categories: the maintenance cost (the cost of servers, networking and power draw) and the deployment cost (the cost of infrastructures)~\cite{edgecost,greenberg2008cost}. The maintenance cost varies temporally and could be possibly affected by many environmental factors in the edge computing scenario, such as the number of service demands and service requirements. The deployment cost, however, is one-shot cost, which makes no difference once the infrastructure has deployed. Thus, in this work, we mainly considered the maintenance cost and did not consider the (fixed) deployment cost in the long-term cost minimization problem. Moreover, the related power or energy cost will be properly ``absorbed" in the components modeled in the service data storage, service computation, and service delivery process.}}

Therefore, during a time period $T$, the servers' cost on cloud nodes can be written as:
\begin{equation}
 \begin{split}
  V^{cloud} &= V_S^{cloud}+V_C^{cloud}+V_M^{cloud} \\
 &= \sum\limits_{u \in U, p \in P, a \in A, t \in T}S_{u,p}(t)\theta_{u,p}^{S,a}(t)V_S \\&+ \sum\limits_{u \in U, p \in P, a \in A, t \in T}C_{u,p}(t)\theta_{u,p}^{C,a}(t)V_C \\& + \sum\limits_{u \in U, p \in P, a \in A, t \in T}(S_{u,p}(t)+S^{'}_{u,p}(t))\theta_{u,p}^{S,a}(t)V_M, 
 \end{split}
\end{equation}
where $V_S^{cloud}$, $V_C^{cloud}$ and $V_M^{cloud}$ are the cost of storage, the cost of computation and the cost of communication in cloud nodes, respectively. $V_S$, $V_C$, and $V_M$ are the cost of per storage unit, the cost of per computation unit and the cost of per communication unit, respectively. The servers' cost on edge nodes is:

\begin{equation}
 \begin{split}
  V^{edge} &= V_S^{edge}+V_C^{edge}+V_M^{edge} \\
 &= \sum\limits_{u \in U, p \in P, e \in E, t \in T}S_{u,p}(t)\alpha_{u,p}^{e}(t)V_{S}^{e} \\&+ \sum\limits_{u \in U, p \in P, e \in E, t \in T}C_{u,p}(t)\beta_{u,p}^{e}(t)V_{C}^e \\&+ \sum\limits_{u \in U, p \in P, e \in E, t \in T}(S_{u,p}(t)+S^{'}_{u,p}(t))\alpha_{u,p}^{e}(t)V_{M}^e,
 \end{split}
\end{equation}
where $V_S^{edge}$, $V_C^{edge}$, $V_M^{edge}$ are the cost of storage, the cost of computation and the cost of communication on edge nodes, respectively. $V_S^e$, $V_C^e$ and $V_M^e$ are the cost per storage unit, the cost per computation unit and the cost per communication unit of a specific edge node $e$, respectively.

\subsubsection*{\textbf{Remark 2}} \emph{The resource price is relatively stable in the current cloud computing market. Thus, we set all cloud nodes with the same storage, computation and communication cost per unit. However, for the edge computing, the resource market is still in an initial and unstable stage, and the resource price of an edge node resources in each EIP is quite different~\cite{Amazon},~\cite{Google}. Therefore, the edge nodes of different EIPs will have different storage, computation, and communication price in our edge federation model.}

Then the total cost of all involved edge servers and cloud servers in an edge federation can be written as:
\begin{equation}
    V=V^{cloud}+V^{edge}.
\label{totalcost}
\end{equation}

The optimization goal is to minimize $V$ over a certain time period. It is worth note that the final optimization result should be subjected to the strict service latency requirements.

\subsection{Guaranteeing the Service Performance}

\emph{1) Modeling the Service Latency}

Latency is the key factor affecting service performance and can be roughly divided into two components, including computing latency and content delivery latency. The computing latency is the time consumption of completing the computation process of services. For an end user $u$, the computing latency of the service $p$ on the cloud and edge servers could be respectively measured by:
\begin{equation}
    l_{u,p}^{cloud,C}(t)=\sum\limits_{a \in A} C_{u,p}(t)\theta_{u,p}^{C,a}(t)\frac{r_{p}}{C_a}, \forall u \in U,\forall p \in P, \forall t \in T,
\end{equation}
\begin{equation}
    l_{u,p}^{edge,C}(t)=\sum\limits_{e \in E}C_{u,p}(t)\beta_{u,p}^e(t)\frac{r_{p}}{C_e}, \forall u \in U,\forall p \in P, \forall t \in T,
\end{equation}
where the parameter $r_{p}$ represents required the computation resource of service $p$ by the end user $u$ and is related to the service category (e.g., social networking, gaming, etc.). Note that, compared to the extra-large computation resources provided by the cloud, the computation resources offered by the edge are limited. Thus, we have $C_a\gg C_e$ in general.

The delivery latency could be divided into the uploading delivery latency and the downloading delivery latency. Users usually access services through a one-hop transmission. Thus, we use the delivery distance instead of the hop distance to estimate the delivery latency in this model. We use $h_u^a$ and $h_u^e$ to denote the delivery distance from cloud node $a$ and edge node $e$ to end user $u$, respectively. First, the service data should be transferred from the user to the server. The uploading delivery latency in the cloud and the edge at time slot $t$ can be estimated as follows, respectively:
\begin{equation}
    l_{u,p}^{cloud,up}(t)=\sum\limits_{a \in A}S_{u,p}(t)\theta_{u,p}^{S,a}(t)h_u^a, \forall u \in U,\forall p \in P, \forall t \in T,
\end{equation}
\begin{equation}
    l_{u,p}^{edge,up}(t)=\sum\limits_{e \in E}S_{u,p}(t)\alpha_{u,p}^e(t)h_u^e, \forall u \in U,\forall p \in P, \forall t \in T.
\end{equation}

Then, after processing in the server, the processed service data will be returned to the users. Thus, the downloading delivery latency in the cloud and the edge at time slot $t$ can be desribed as:

\begin{equation}
    l_{u,p}^{cloud,do}(t)=\sum\limits_{a \in A} S_{u,p}^{'}(t)\theta_{u,p}^{S,a}(t)h_u^a, \forall u \in U,\forall p \in P, \forall t \in T,
\end{equation}
\begin{equation}
    l_{u,p}^{edge,do}(t)=\sum\limits_{e \in E}S_{u,p}^{'}(t)\alpha_{u,p}^e(t)h_u^e, \forall u \in U,\forall p \in P, \forall t \in T.
\end{equation}

\emph{2) The Constraint on the Service Latency}

The service demands of services usually vary temporally and spatially for heterogeneous end users. Hence, we should make sure that the required performance of services (e.g., latency requirement) can be guaranteed by the schedule of edge federation. Let $l_p$ denote the required latency of accessing service $p$. In any time slot $t$, only when the actual latency does not exceed $l_p$, the service can be regarded as satisfied in that time slot. Therefore, the relationship between the actual latency and required latency can be defined as:

\begin{equation}
 \begin{split}
  l_{u,p}(t) &= l_{u,p}^{cloud}(t)+l_{u,p}^{edge}(t) \\
 &= [l_{u,p}^{cloud,S}(t)+l_{u,p}^{cloud,C}(t)]+[l_{u,p}^{edge,S}(t)+l_{u,p}^{edge,C}(t)] \\
 &\leq l_p,
 \end{split}
\end{equation}
where $l_{u,p}(t)$ denotes the actual latency of service $p$ from end user $u$ at time slot $t$. Then, we use a satisfaction parameter $m_{u,p}(t)$ to represent whether a service demand of the user $u$ is guaranteed, which can be defined as:

\begin{equation}
m_{u,p}(t)=
\left\{
             \begin{array}{lr}
             1 \quad,\quad l_{u,p}(t)\leq{l_p}, & \\
             0 \quad, \quad l_{u,p}(t)>{l_p}.
             \end{array}
\right.
\end{equation}

Moreover, edge federation needs to keep the corresponding services at a high-level performance in the business environment to attract more users and improve revenues. The overall performance of service $p$ in edge federation can be measured by the satisfaction ratio $r_p^{sat}$, which can be written as:
\begin{equation}
    r_{p}^{sat}= \frac{\sum\limits_{t \in T}\sum\limits_{u \in U}{S_{u,p}(t)m_{u,p}(t)}}{\sum\limits_{t \in T}\sum\limits_{u \in U}{S_{u,p}(t)}}.
\end{equation}

According to the existing industry standards, the satisfaction ratio should reach following range:
\begin{equation}
    l_1\leq r_{p}^{sat}\leq l_2,
\label{constraint7}
\end{equation}
where $l_2$ could be 100\%, and $l_1$ is usually larger than 99\%.

The satisfaction ratio of service $p$ is evaluated by the satisfied service demands of each user in every time slot. Thus, we accumulate those service demands, whose latency requirements have been satisfied, to calculate the satisfaction ratio of a specific service. Note that calculating the service satisfaction with a global measure is inaccurate, such as the average service latency, due to the potential uneven distribution (e.g., a bimodal distribution) of the service latency for each user.

Therefore, under the latency constraints, the problem that the central optimizer needs to solve for the edge federation can be formulated as the following optimization problem:

\begin{subequations} \label{eq:optimization}
    \begin{align}
        &\underset{\{\theta_{u,p}^{S,a}(t), \theta_{u,p}^{C,a}(t), \alpha_{u,p}^e(t), \beta_{u,p}^e(t)\}}\min\quad V \label{eq:optimization-a}\\
        &\text{s.t.}\quad(\ref{constraint1})\sim(\ref{constraint4}),(\ref{constraint5})\sim(\ref{constraint6})\quad\text{and}\quad(\ref{constraint7}). \label{eq:optimization-b}
    \end{align}
\end{subequations}

By solving this optimization problem, we can find the optimal resource assignment schedules (e.g. optimal caching and computing variables) of edge federation at every time slot. Thus, edge federation can achieve following superiorities:

\textbf{Scalability}: although the edge resources of EIPs are limited, the cloud resource could be the important supplement (vertical integration) and enable EIPs to easily serve the service demands with elastic resource configuration. For instance, if there are huge-amount service demands beyond the capacity of edge resources, edge federation can improve $\theta_{u,p}^{S,a}(t)$ and $\theta_{u,p}^{C,a}(t)$ to utilize more cloud resources. Then, as pointed in (17) and (18), the $\alpha_{u,p}^{e}(t)$ and $\beta_{u,p}^{e}(t)$ will be reduced. With such the adjustment, EIPs could enhance their resource capacities and push services with low latency requirements to the cloud and thus leave more edge resources for the coming services with high latency requirements.

\textbf{Efficiency}: as pointed in (28) to (31), the service latency requirements should be guaranteed in the service provisioning process. Therefore, with the latency constraints, the central optimizer of edge federation gives the optimal service provisioning schedule to minimize the cost of EIPs. Note that, with the target of minimizing the cost of EIPs, the service latency is unnecessary as low as possible, but rather to control the latency (i.e., the variables $\theta_{u,p}^{S,a}(t)$, $\theta_{u,p}^{C,a}(t)$, $\alpha_{u,p}^{e}(t)$ and $\beta_{u,p}^{e}(t)$ ) just satisfy the requirement, and then EIPs could use the cheap cloud resources as more as possible to avoid the high edge overhead. Therefore, edge federation can always achieve efficient and qualified service delivery under different service requirements. 

\textbf{Low Latency}: due to the horizontal resource integration, edge federation could have more edge nodes to deploy service in a wide geographical area. Edge federation is able to place services closer (e.g., the smaller $h_u^e$ in (25) and (27)) to the end user by adjusting the edge resource variables (i.e., $\alpha_{u,p}^{e}(t)$ and $\beta_{u,p}^{e}(t)$). Although in this paper, we mainly focus on the cost minimization problem instead of minimizing the service latency, the above operation could possibly make better use of the edge resources and enable the lower service latency (i.e., the smaller accumulated delivery distance $\sum\limits_{e \in E}h_u^e$ in (25) and (27)).

\section{Problem Transformation and Dynamic Resolving Algorithm}\label{sec:algorithm}

In this section, we propose a dimension-shrinking method to reformulate the optimization problem into an easily solved form. Based on this method, we further develop a dynamic service provisioning algorithm to deal with varying service demands.

\subsection{Problem Transformation}

In our problem, variables $\theta_{u,p}^{S,a}(t)$, $\theta_{u,p}^{C,a}(t)$, $\alpha_{u,p}^{e}(t)$ and $\beta_{u,p}^{e}(t)$ are related to four factors including the edge nodes/cloud nodes, end users, services and time slots. Hence, when we formulate the optimization problem of each time slot, we find that the variable matrix is the four-dimension matrix, which is hard to solve with the existing solvers and can be time-consuming. Therefore, we reformulate this problem as a low-dimension optimization problem so that it can be solved efficiently using the off-the-shelf solvers. Particularly, in our transformation, we transform these variable matrices to the two-dimension matrices.

We use $V_{S}^{edge}$, part of the $V$, as an example to illustrate the transformation process. To ease understanding, we begin with a simple scenario where only one service and a single time slot (i.e., $|P|$=1, $|T|$=1) are considered. Then, the original four-dimension caching variables can be converted to two-dimension variables, e.g., $\alpha_{u,p}^{e}(t)$ can be coverted to $\alpha_{u}^{e}$, where ${u}\in{U}$, ${e}\in{E}$. Assume that $|U|=i$, $|E|=j$, the variable matrix of $\alpha_{u}^{e}$ can be written as:
\begin{equation}
\bm{\alpha} =
\left[
\begin{matrix}
\alpha_1^1&\alpha_1^2&\cdots&\alpha_1^j\\
\alpha_2^1&\alpha_2^2&\cdots&\alpha_2^j\\
\vdots&\vdots&\ddots&\vdots\\
\alpha_i^1&\alpha_i^2&\cdots&\alpha_i^j&
\end{matrix}
\right],
\label{matrix1}
\end{equation}
where each $\alpha_{i}^{j}$ means the fraction of storage demands requested by user $u$ and assigned to edge node $e$. Let the vector $\bm{S}= ( S_1, S_2, \cdots, S_i)^T$ denote the amount of storage demands from each of these end user. The vector $\bm{V_S^{E}}= (V_S^{e_1}, V_S^{e_2}, \cdots, V_S^{e_j})$ represent the cost of per storage unit in different edge nodes. Therefore, the $V_{S}^{edge}$ part can be formulated as:

\begin{equation}
   V_{S}^{edge} = \lVert(\bm{S}\bm{V_S^{E}})\circ\bm{\alpha})\rVert_1.
\end{equation}
where the symbol "$\circ$" denotes the Hadamard product of two matrices, and each element of the matrix $(\bm{S}\bm{V_S^{E}})\circ\bm{\alpha}$ represents the cost at a certain edge node.

So far, we consider the solution for a more general case that includes multiple services and multiple time slots (i.e., $|P|=m$ and $|T|=n$). In this case, the martrix of variable $\alpha_{u,p}^{e}(t)$ can be converted into a super matrix that consists of $m*n$ aforementioned two-dimension matrices (refer to (\ref{matrix1})). Thus, the variable $\alpha_{u,p}^{e}(t)$ could be extended as following:

\begin{equation}
   \bm{\widehat{\alpha}} = [\bm{\alpha}{(1)},\bm{\alpha}{(2)},\cdots,\bm{\alpha}{(m*n)}]^T,
\end{equation}
where each matrix $\alpha(l)$ represents the matrix of storage variable $\alpha_{u}^{e}$ of a certain service $\hat{p}$ at time slot $t$, and $m*(t-1)+\hat{p}=l$.

The vector $\bm{S}$ and $\bm{V_S^{E}}$ could also be extended for the general case as follows:
\begin{equation}
   \bm{\widehat{S}} = [\bm{S}{(1)},\bm{S}{(2)},\cdots,\bm{S}{(m*n)}]^T,
\end{equation}

\begin{equation}
   \bm{\widehat{V_S^{E}}} = [\bm{V_S^{E}}{(1)},\bm{V_S^{E}}{(2)},\cdots,\bm{V_S^{E}}{(m*n)}].
\end{equation}

In both of which each $\bm{S}(l)$ and $\bm{V_S^{E}}{(l)}$ represent the storage demand vector and the edge caching cost vector of service $\hat{p}$ at time slot $t$, respectively. Thus, $m*(t-1)+\hat{p}=l$. The $V_S^{edge}$ could be converted to:

\begin{equation}
   V_S^{edge} = \lVert(\bm{\widehat{S}}\bm{\widehat{V_S^{E}})\circ\bm{\widehat{\alpha}})}\rVert_1.
\end{equation}
In this way, we reduce the variable matrices of $\theta_{u,p}^{S,a}(t)$, $\theta_{u,p}^{C,a}(t)$, $\alpha_{u,p}^{e}(t)$ and $\beta_{u,p}^{e}(t)$ to two dimensions and without any approximation. Thus, there is no loss with the transformation, and the optimization result is still optimal. Finally, the problem (\ref{eq:optimization}) could be solved efficiently with existing LP solvers such as CVX Gurobi solver~\cite{Gurobi}.

\begin{figure*}[t]
\centering
\includegraphics[width=16cm]{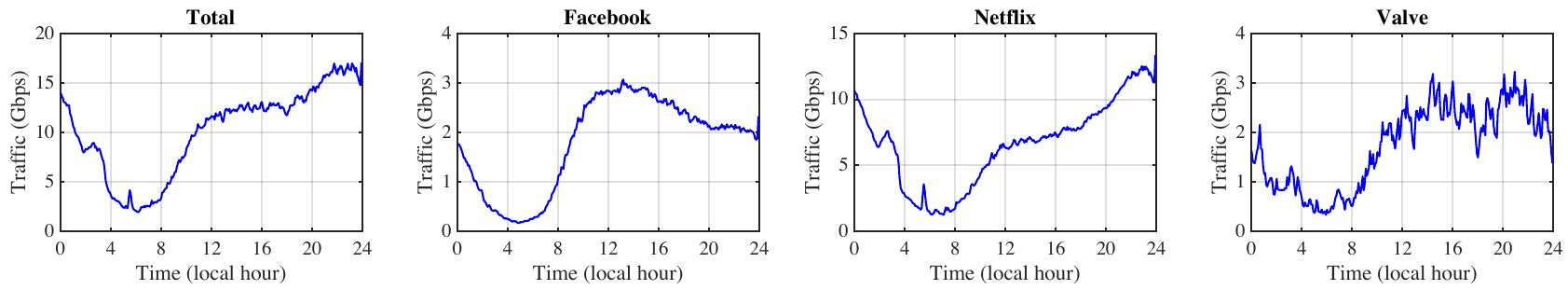}
\caption{Traffic demands of different services at NORDUnet nodes on May. 07, 2017.}
\label{fig:dailytraffic}
\vspace{-0.2in}
\end{figure*}

\begin{figure}[t]
\centering
\includegraphics[width=9cm]{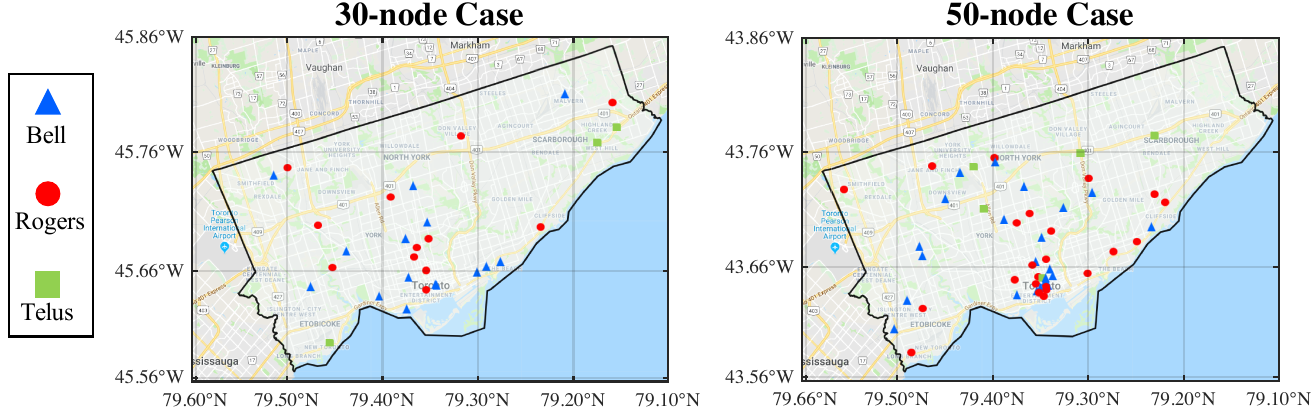}
\caption{The base station map of Toronto city.}
\label{fig:toronto_map}
\vspace{-0.1in}
\end{figure}

\subsection{Service Provisioning Algorithm}
After the transformation mentioned above, we further develop a dynamic resolving algorithm, named \emph{SEE} (Service provision for Edge fEderation), to achieve an efficient service provisioning solution in the edge federation environment.

To be more specific, as shown in \textbf{Algorithm 1}, our algorithm is developed under the dynamic service demands; thus the service provisioning should be rescheduled at each time slot. We take the storage and computation capacities of cloud nodes and edge nodes ($C_a$, $S_a$, $C_e$ and $S_e$), the profile of services' computation and latency requirements (i.e., $r_p$ and $l_p$), and the transmission distance ($h_e$ and $h_a$) as the inputs of our algorithm. In each time slot, we first predict the demands of services by a well-studied method (e.g., ARIMA). 
Based on the prediction results, the edge federation could solve the optimization problem (\ref{eq:optimization}) and calculate the schedule for the next time slot in advance. Such an optimization process is mainly executed by the consortium of edge federation for enabling dynamic optimal service provisioning. It decides how much workload retain at the edge or offload to the cloud, and how to deploy services among heterogeneous edge servers and cloud servers.

\begin{algorithm}[t]
        \caption{\emph{SEE} algorithm}
        \begin{algorithmic}[1]
            \REQUIRE: $C_a$, $S_a$, $C_e$, $S_e$, $r_p$, $l_p$, $h_e$, $h_a$
            \ENSURE: edge storage variable $\alpha^e_{u,p}(t)$ d, edge computation variable $\beta^e_{u,p}(t)$, cloud storage variable $\theta^{S,a}_{u,p}(t)$, and cloud computation variable $\theta^{C,a}_{u,p}(t)$;
            \FOR{$t_1$ to $t_n$}
                \STATE{Predict the service demands of different services $K_{u,p}(t)$=($S_{u,p}(t_i)$, $S_{u,p}^{'}(t_i)$, $C_{u,p}(t_i)$})
                \STATE{Update the $\alpha^e_{u,p}(t_i)$, $\beta^e_{u,p}(t_i)$, $\theta^{S,a}_{u,p}(t_i)$, $\theta^{C,a}_{u,p}(t_i)$ by solving the optimization problem (32a) }
                \STATE{Calculate the cost of EIPs at time slot $t_i$: $V(t_i)=(V^{edge}(t_i)+V^{cloud}(t_i))$}
            \ENDFOR
        \end{algorithmic}
\end{algorithm}

\section{Experimental Evaluation}\label{sec:eva}
In this section, we conduct trace-driven experiments over the base station network in Toronto, Canada and evaluate the performance of our service provisioning model under a multi-EIP network environment. We measure the performance of the edge federation in terms of the total cost of EIPs for serving a given set of edge services.

\subsection{Experimental Settings}

\emph{1) Designed Network}

We first obtain the datasets of the edge-computing environment from the published data of the Canada government~\cite{canadadata}, which provides the details of the location and company of base stations all over Canada. We make use of the base station dataset for the following reasons: i) upgrading base stations to edge nodes is a reasonable and accessible solution for the construction of future edge-computing environment; ii) the datasets have specific location information of base stations. Hence, based on the fact that round trip time between two nodes is approximately linear with their geo-distance~\cite{krajsa2011rtt}, the content delivery latency could be accurately estimated.

The designed network of our experiments is constructed across the region of the Toronto city. As shown in Fig.~\ref{fig:toronto_map}, we carefully select the amount of 30 and 50 base stations as the potential edge nodes by fully considered the density of the population 
and the business condition in different areas of the city. In common sense, compared with the non-flourishing area, the larger number of edge nodes are needed in the prosperous and populous area. All of the base stations are chosen from three popular telecommunication providers in Canada, including the  Bells, Rogers, and Telus. Then, we further select the Amazon Datacenter in Montreal, the Google datacenters in the United States as the potential cloud nodes in our experiments.\footnote{Under the common situation, most of the users in the world are hundreds of kilometers away from the data center, and some of them even need to get the service from the data center continental distance away~\cite{Googlelocation}. To make our experiment more representative, we carefully select the data centers far away from Toronto city as the potential cloud nodes.} Therefore, these selected edge nodes and cloud nodes make up the designed network in this paper.

\begin{figure*}[t]
\centering
\subfigure[Overall service provisioning cost of all EIPs under the 7 latency requirements.]{\label{fig:total_cost}\includegraphics[width=0.49\textwidth]{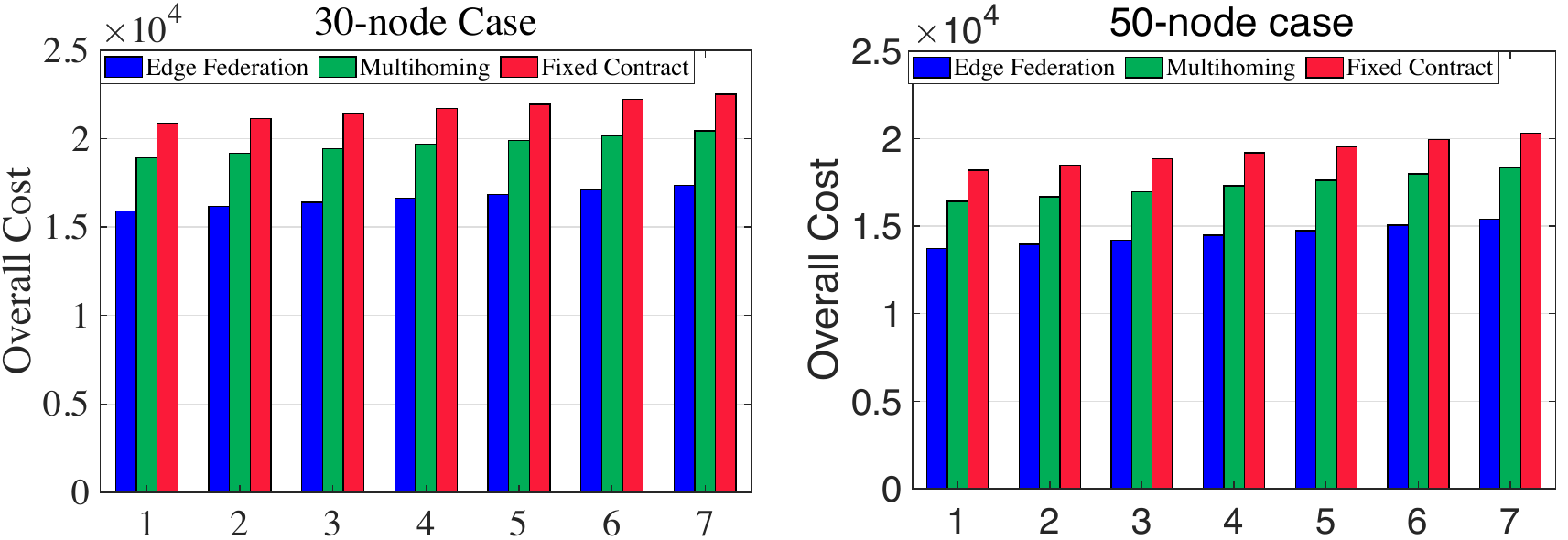}}
\subfigure[Cost savings of edge federation with varying latency requirements over the 30-node case and 50-node case, repectively.]{\label{fig:boxplot}\includegraphics[width=0.49\textwidth]{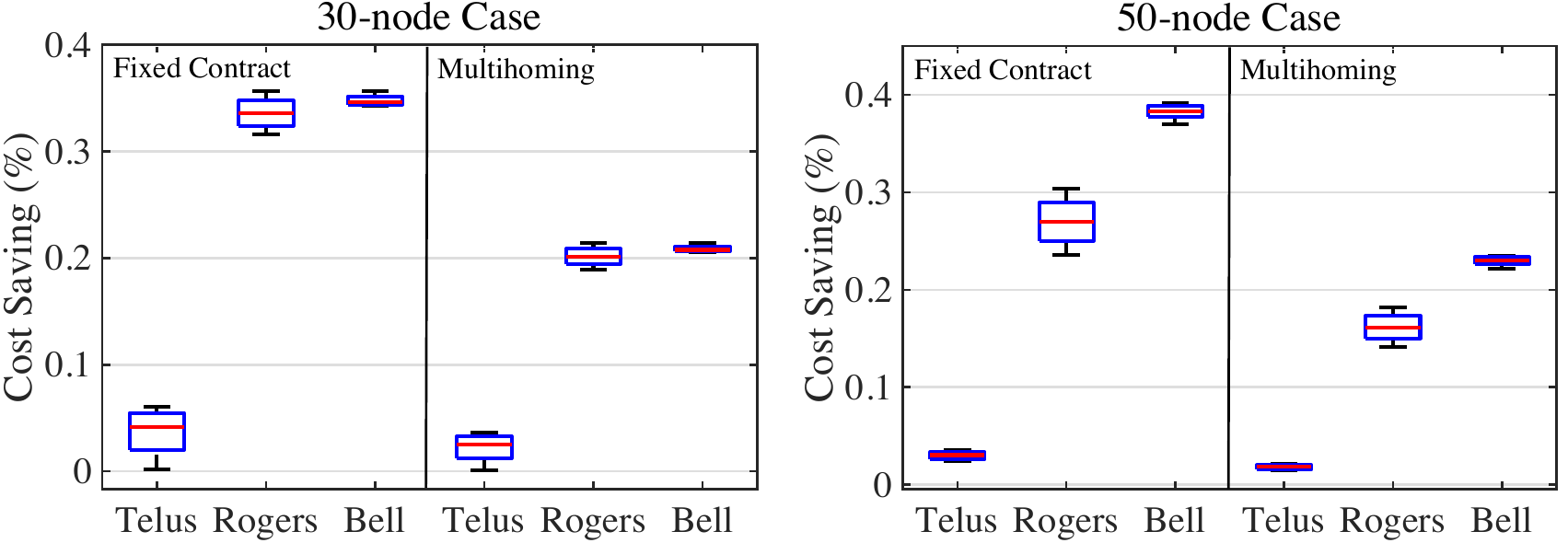}}
\caption{Overall service provisioning performance of three EIPs in the fixed contract model, multihoming model and edge federation model.}
\label{fig:the overall cost}
\vspace{-0.1in}
\end{figure*}

\emph{2) Service Demands of End Users}

Then, we collect the service traffic data from the NORDUnet\footnote{http://stats.nordu.net/connections.html}, a research and education oriented network infrastructure that hosts caching servers at various peering points over Europe and North America. By using these real-world trace data, we generate synthetic service demands of end users at each location in our designed network.

\textbf{Dynamic Service Demands}: We mainly consider three types of services in our paper, including online gaming, online video, and social media. They represent the high, normal, and low latency requirements, respectively. Thus, we correspondingly select three representative services, including Valve, Netflix, and Facebook. Fig.~\ref{fig:dailytraffic} shows the traffic curves in a 24-hour time window on May. 7, 2017. 
There are some interesting observations of the service traffic: Netflix accounts for the most significant portion of the traffic. The peak demands of Valve and Netflix appear at night, while the peak demands of Facebook appear in the daytime.

\textbf{Synthetic Traffic Generation}: Referring to the above service demand patterns, we generate synthetic traffic demands for the evaluation. First, we normalize the traffic demand of each service as the traffic profile, i.e., $q_p(t), \forall{t} \in T$. Then, we collect the amount and density of the population of Toronto, from the online published data~\cite{population},~\cite{density}. Based on that, we can generate the synthetic service demands for the location area of each end user by calculating (\ref{generate 1})-(\ref{generate 3}) of different types of services. This traffic information is treated as the result of \emph{Traffic Analyser} in Fig.~\ref{fig:architecture-of-EF} and sent to the \emph{Central Optimizer} for calculating the optimal service provisioning and requesting schedules.

\begin{center}
\centering
\begin{table}
\caption{Latency requirements for three services.}
\label{table_requirement}
\begin{tabular}{ | p{1.6cm} | p{0.5cm}  | p{0.5cm}  | p{0.5cm}  | p{0.5cm}  | p{0.5cm}  | p{0.5cm}  | p{0.5cm}  |}

\hline

\diagbox[width=5em,trim=l]{Service}{Group} & 1 & 2 & 3 & 4 & 5 & 6 & 7 \\ \hline\hline
Facebook & 72 & 68 & 64 & 60 & 56 & 52 & 48 \\ \hline
Valve & 36 & 34 & 32 & 30 & 28 & 26 & 24 \\ \hline
Netflix & 54 & 51 & 48 & 45 & 42 & 39 & 36 \\ \hline

\end{tabular}
\vspace{-0.2in}
\end{table}
\end{center}

\vspace{-0.32in}
\subsection{Performance Evaluation}
In this part, we analyze the service provisioning process of Telus, Rogers, and Bell in both the 30-node case and 50-node case. We mainly refer to two service provisioning models to compare with ours in edge federation. 1) Fixed contract model: each ESP can only contract with one EIP. To test the performance of the fixed contract model, we assume several fixed relationships: Telus contracts with Facebook, Rogers contracts with Valve, and Bell contracts with Netflix. 2) Multihoming: one ESP can contract with several EIPs, where each EIP manages its resource independently without the global view. We assume that Telus contracts with Facebook and Valve, Rogers contracts with Valve and Netflix and Bell contracts with Netflix and Facebook. Moreover, to achieve fairness, the computation and storage capacity of edge and cloud nodes are set to be the same in different models.
We evaluate the performance of the existing service provisioning models and our edge federation model by the cost of EIPs, under different latency requirements of services and the varying amount of service demands. Particularly, we consider not only the total cost but also the average cost, which could be defined as follow:

\begin{itemize}
    \item{The total cost: the overall cost in total 24 time slots for all EIPs, which can be calculated by (\ref{totalcost}).}
    \item{The average cost: the average cost of each EIP for end users at each location at time slot $t$ can be defined as
    \begin{equation*}
    \begin{aligned}
    v_{u,p}(t)=&[\sum\limits_{u \in U}\sum\limits_{e \in E}S_{u,p}(t)\alpha^{e}_{u,p}(t)V_S^e+\\
    &\sum\limits_{u \in U}\sum\limits_{e \in E}S_{u,p}(t)\beta^{e}_{u,p}(t)V_C^e)]/n_{users},
    \end{aligned}
    \end{equation*}
    where $n_{users}$ represents the number of users.}
\end{itemize}

\begin{figure*}[t!]
  \centering
  \subfigure[Cost saving of three EIPs.]{\label{fig:average_cost}\includegraphics[width=0.24\textwidth]{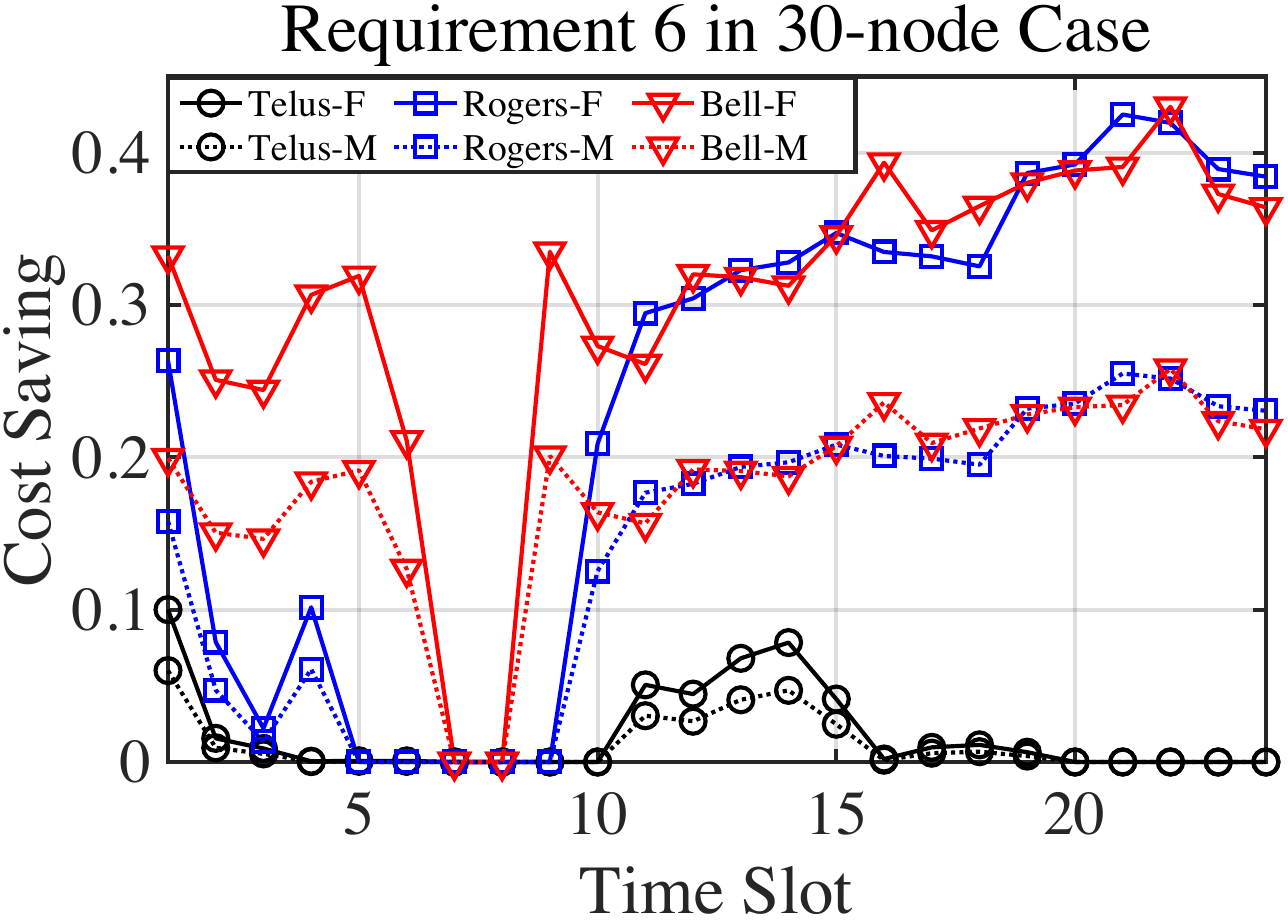}}
  \subfigure[Average cost of Rogers.]{\label{fig:rogers}\includegraphics[width=0.24\textwidth]{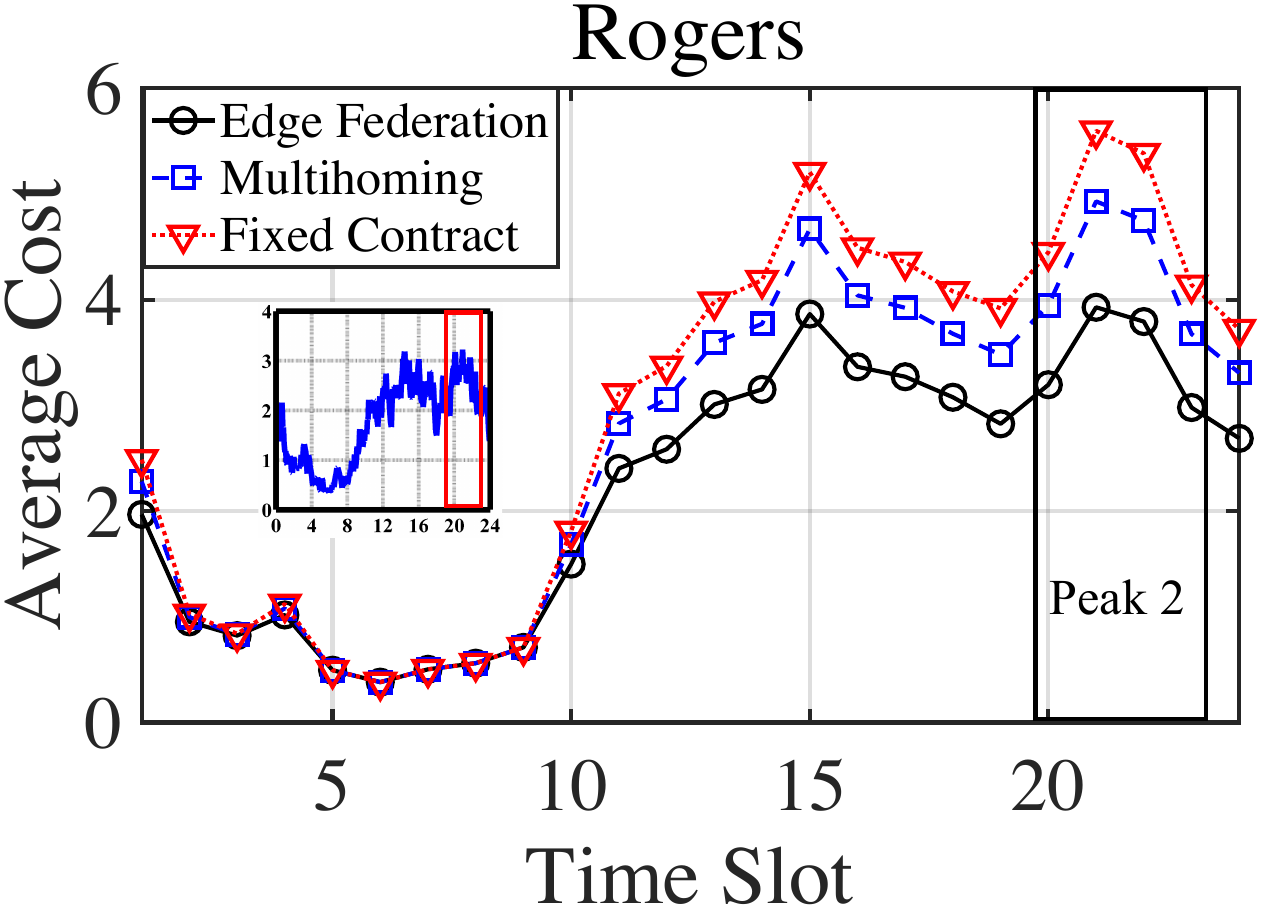}}
  \subfigure[Average cost of Bell.]{\label{fig:bell}\includegraphics[width=0.24\textwidth]{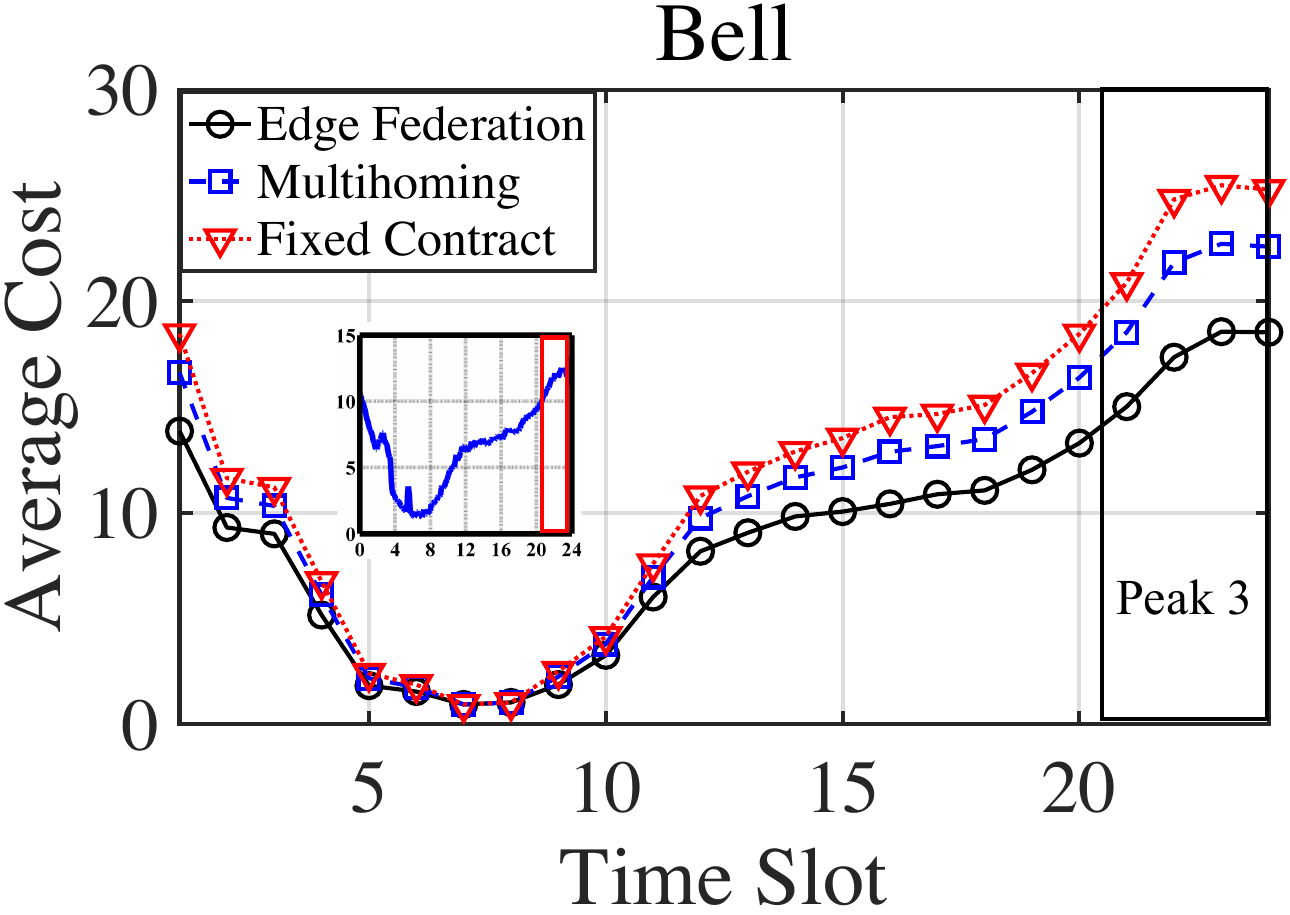}}
  \subfigure[Average cost of Telus.]{\label{fig:telus}\includegraphics[width=0.24\textwidth]{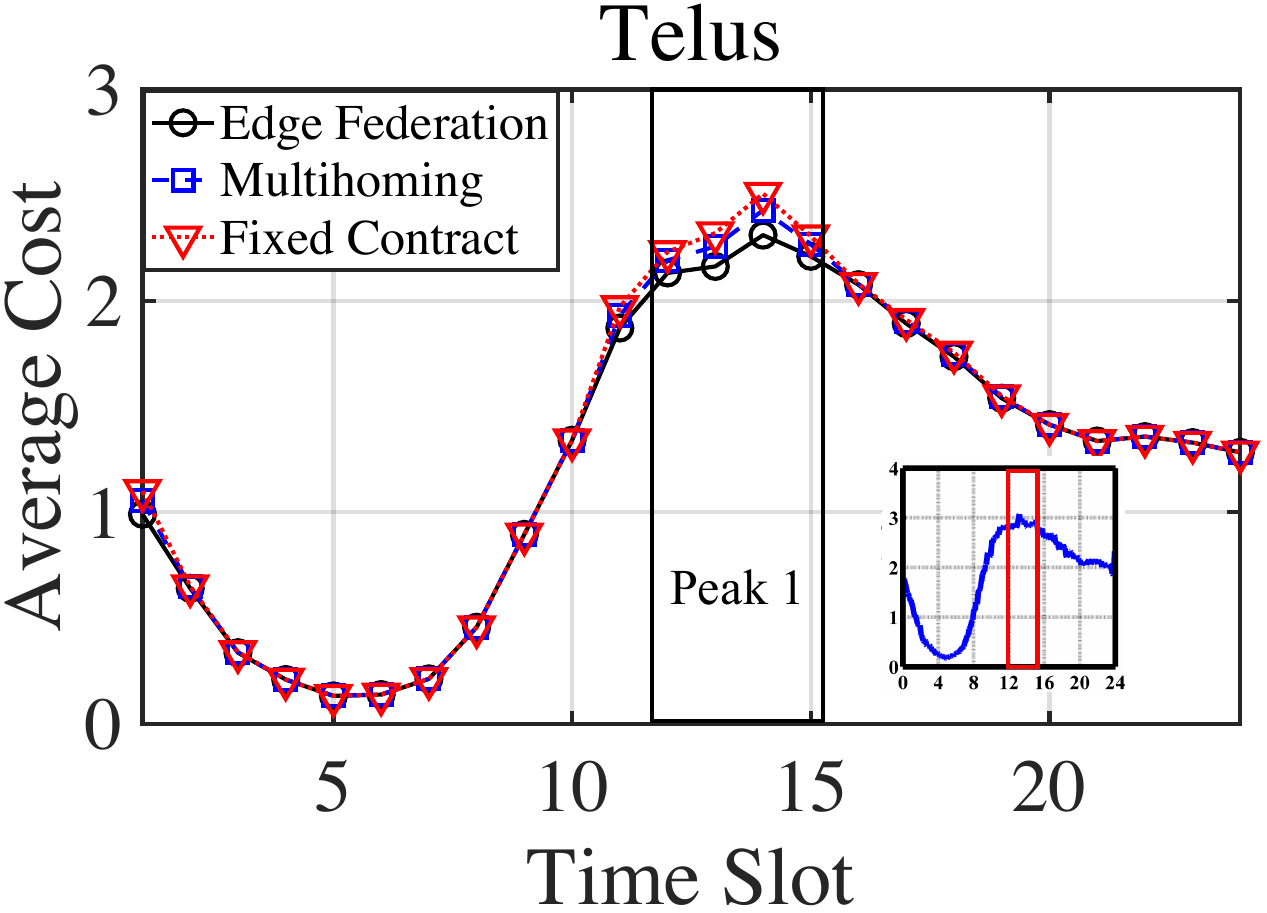}}
  \caption{Service provisioning cost of three EIPs, under the fixed contract, multihoming and edge federation model in 30-node case. In Fig. \ref{fig:average_cost}, F and M represent the fixed contract and multihoming model, respectively.}
  \label{average_cost_temporal}
\vspace{-0.2in}
\end{figure*}

\emph{1) The Overall Performance comparison}

Fig.~\ref{fig:total_cost} presents the overall cost of the 50-node and 30-node cases with various latency requirements. Detailed requirements are shown in Table.~\ref{table_requirement}, where the smaller number means the more strict requirement. First, we can observe that, with the requirements from loose to strict, the overall cost increases. This is consistent with our expectations that the edge resource is used to provide low service latency, higher requirements of latency will incur more usage of edge resources (than that of the cloud). Because the edge resource has a higher cost per unit than the cloud resource, the overall cost increases. Moreover, it can be seen that compared with other models, the edge federation can be more cost-efficient for EIPs and achieve better service provisioning performance. For instance, compared with the multihoming model and fixed contract model, edge federation can achieve average 15.5\% and 23.3\% savings in 30-node case, and average 16.3\% and 24.5\% savings in 50-node case, respectively. The saving will be significant for EIPs, especially for those with a large number of edge nodes in the extensive coverage area. There is also an interesting phenomenon that in each latency requirement group, the total cost of the 50-node case is lower than the cost in the 30-node case. This could be concluded that: compared with the 30-node case, the 50-node case has more and better (i.e., shorter distance to the end user) options in a specific area for the EIPs' deployment. Thus, EIPs can avoid remote service deployment, and the cost of service delivery could be saved.

\subsubsection*{\textbf{Does Service Type Matter?}}
To figure out whether or not the type of service has a significant impact on the cost saving, we analyze the performance of each EIP individually under the constraints of varying latency requirements from latency requirement groups 1 to 7 in Table.~\ref{table_requirement}. The corresponding results are shown in Fig.~\ref{fig:boxplot}, where the range of cost savings for each EIP is given. We can see that, compared with the fixed contract model, Telus, Rogers, and Bell save about 3.7\% and 3.0\%, 33.6\% and 26.0\%, 34.8\% and 38.1\% of the costs in 30-node and 50-node cases, respectively; compared with the multihoming model, Telus, Rogers, and Bell save about 2.5\% and 1.5\%, 20.1\% and 16.1\%, 20.6\% and 22.2\% of the costs in 30-node and 50-node cases, respectively. Such results indicate that the edge federation is advantageous for all kinds of EIPs, irrespective of the services they contracted and the number of nodes they have.

In addition, we find that the EIP contracted with the higher latency requirement service will receive even greater cost saving.\footnote{In this paper, we assume that the latency requirement of the Valve is more strict than the Netflix, and the Netflix is more strict than Facebook.} The reason for this result may be due to the fact that higher latency requirement services need to use more edge resources. However, the limited-capacity and low-coverage edge resources in an individual EIP make it difficult for EIPs deploying and provisioning service in an efficient way, i.e., due to the considerable accumulated distance between different edge nodes, the service provisioning process will cause significant service delivery cost. Such a dilemma can be significantly alleviated by the resource integration in the edge federation, and thus the cost of service delivery could be reduced, especially for the high latency requirement services.

\begin{figure*}[t!]
  \centering
  
  \subfigure[EIPs' cost saving over seven different latency requirements.]{\label{fig:robustness}\includegraphics[width=0.245\textwidth]{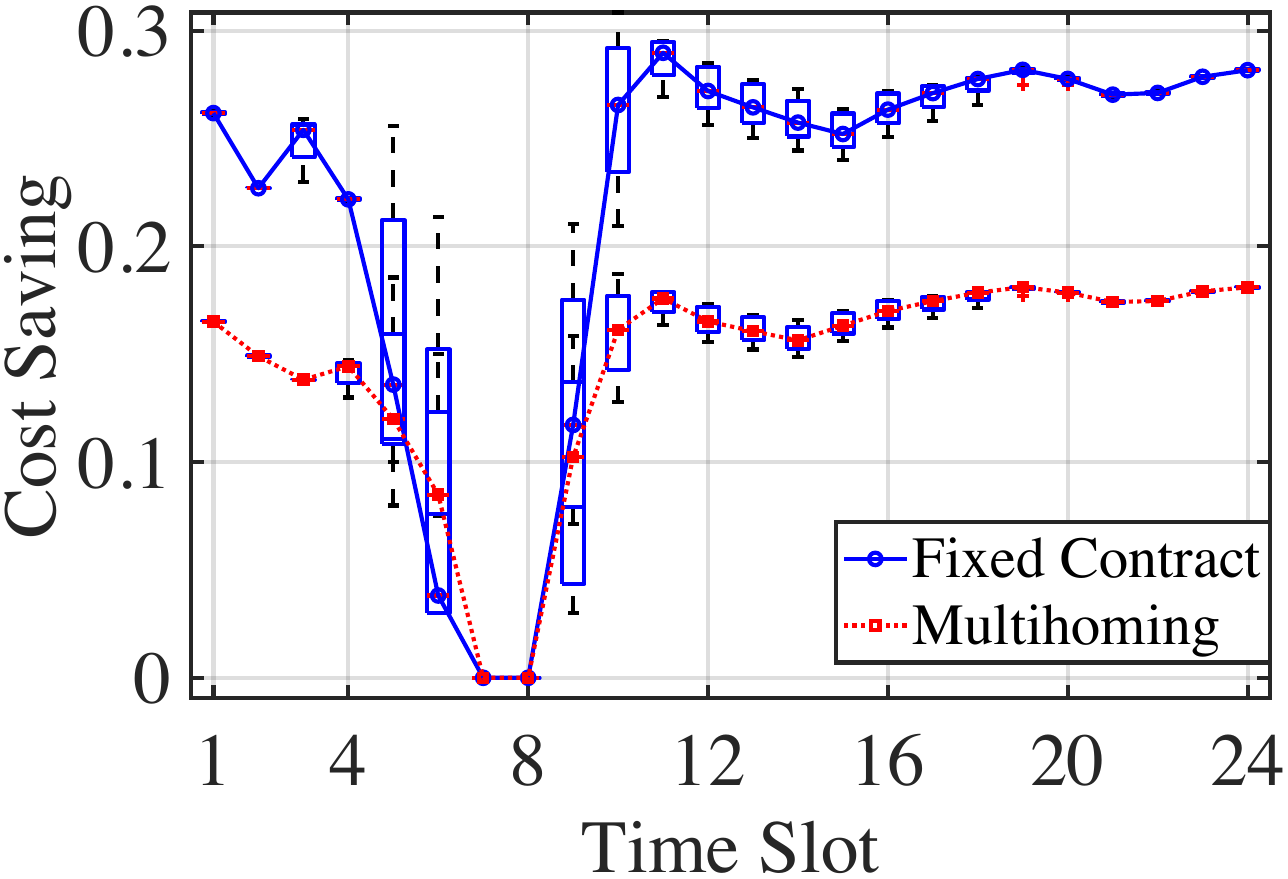}}
  \subfigure[Validation of the horizontal extending edge nodes.]{\label{fig:extending}\includegraphics[width=0.235\textwidth]{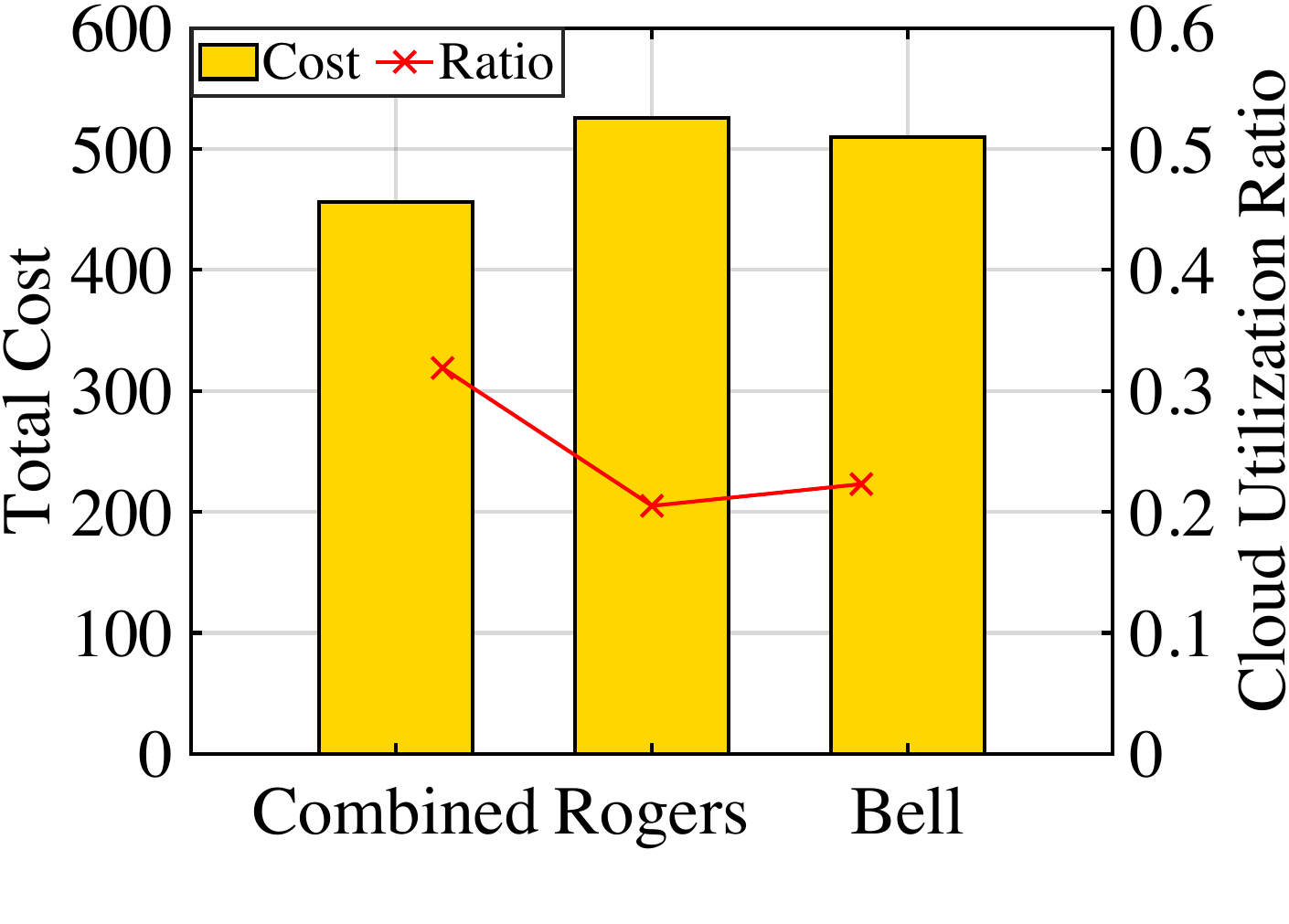}}
  \subfigure[Edge and cloud resource utilization ratio in different ESPs.]{\label{fig:utilization}\includegraphics[width=0.24\textwidth]{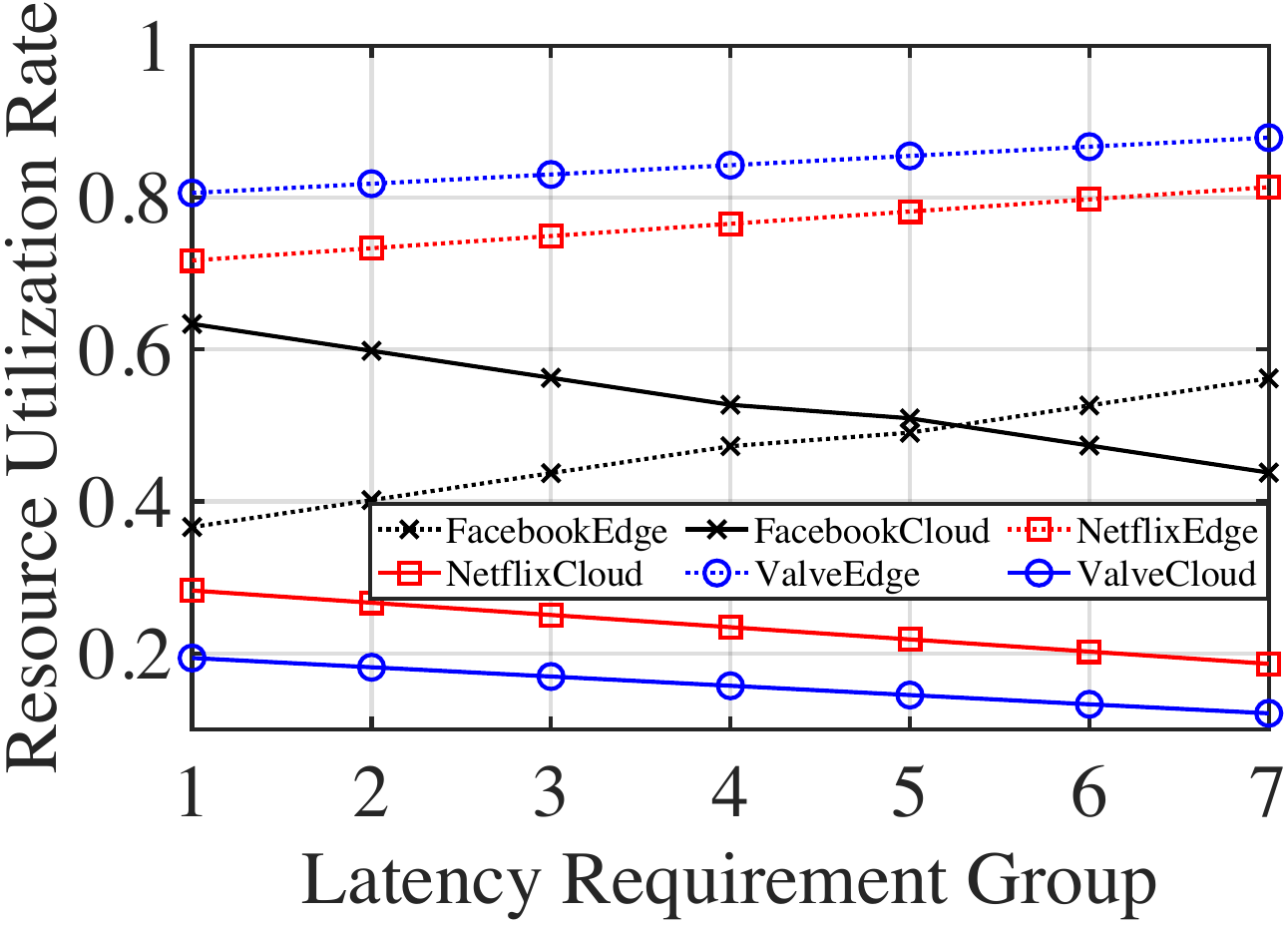}}
  \subfigure[EIPs' cost saving over different lengths of time slot.]{\label{fig:diff_timeslot}\includegraphics[width=0.245\textwidth]{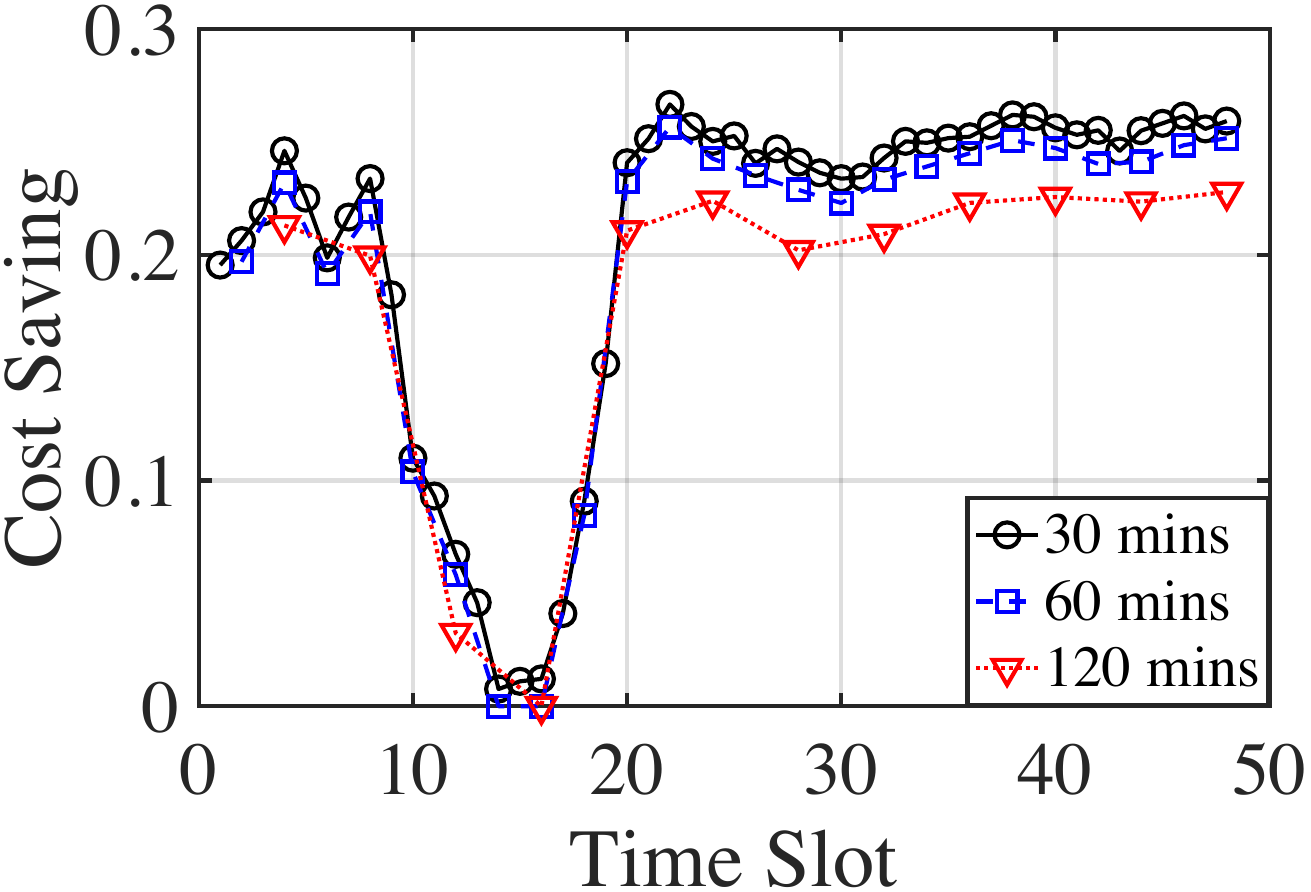}}
  \caption{Comparison of the service provisioning cost and the resource utilization ratio.}
\vspace{-0.25in}
\end{figure*}

\emph{2) Variance of the Cost Saving over Dynamic Traffic}

In the previous part, we accumulate the total cost over the whole period without considering the cost variation of each time slot. In this part, given the fixed latency constraints of the services (i.e., group six in Table.~\ref{table_requirement}), we consider the cost variation under dynamic service demands (i.e., from the time dimension). We show the average cost of Rogers, Bell and Telus over the whole time period in Fig.~\ref{fig:rogers},~\ref{fig:bell} and~\ref{fig:telus}, respectively, and have two major observations.
First, combined with Fig.~\ref{fig:dailytraffic} (i.e., the inset plot in Fig.~\ref{fig:rogers},~\ref{fig:bell} and~\ref{fig:telus} ), we can easily find that the average cost curve of edge federation oscillates as time varies and has similar changing trend with the amount of the service demand, e.g., the peak of the average cost curve is consistent with that of the service demand curve.
Second, regardless of the service demand variations, our edge federation model always outperforms the other two models in both 30-node and 50-node cases, with 11.8\% and 17.6\% cost savings for Rogers, 15.1\% and 22.6\% for Bell, and 1.3\% and 1.8\% for Telus, respectively.

\subsubsection*{\textbf{Does Amount of Service Demands Matter?}}
Figure~\ref{fig:average_cost} shows the cost saving of three EIPs under the latency requirement 6 in 30-node case. Combined with Fig.~\ref{fig:dailytraffic}, there is a similar changing trend between the amount of service demands and cost savings, and the higher cost saving is likely to occur when service demand is larger, which demonstrates that the cost saving has a strong correlation with the amount of service demands. For instance, we use a 4-time-slot window to circle the peak cost of each service in Fig.~\ref{average_cost_temporal}. \emph{Peak 1} (time slot 12 to 15), \emph{Peak 2} (time slot 20 to 23) and \emph{Peak 3} (time slot 21 to 24) represent the peak costs of Telus, Rogers and Bell, respectively. It is clear that the time windows of peak cost saving are perfectly matched with the peak service demands in each service. This means that the edge federation achieves even better performance in the case of the larger amount of service demands. It could be significantly helpful in the practical huge-traffic network environment.

\emph{3) Strength of Edge Federation}

\subsubsection*{\textbf{Resilient and Robust Service Provisioning}} \emph{Can edge federation achieve good performance all the time under varying requirements and dynamic service demands?} The question is critical to justify whether the edge federation can be reliable to the real network environment. 

As shown in Fig.~\ref{fig:robustness}, to answer this question, we mainly analyze the performance from both the time dimension and the latency requirement dimension in 30-node case. From the time dimension, compared with the fixed contract model and multihoming model, we can easily observe that the cost savings are positive all the time, which means edge federation outperforms other models no matter how much service demands are required. From the dimension of varying requirements, the edge federation shows steady cost savings with minor fluctuations over different latency requirements. There is an interesting phenomenon: the cost saving has the relative big fluctuations from time slot 5 to time slot 10, whereas the performance oscillates within a small range in the other time. 
To figure out the underlying rationale, we then check the daily traffic in Fig.~\ref{fig:dailytraffic} and find that the traffic from the time slot 5 to the time slot 10 is much lower than the other time slots. This indicates that edge federation could achieve more stable performance in the massive traffic scenario than in the light traffic scenario. This result once again proves that edge federation is suitable for the real huge-traffic environment.

\subsubsection*{\textbf{Cost-Efficiency Function with Horizontal Extending Edge Nodes}} Edge federation enables the horizontal extension by integrating the edge nodes of EIPs. \emph{Is this extending edge nodes function can indeed reduce the cost of EIPs?} For validating the effectiveness of the horizontal extending, we specially select two EIPs: Rogers and Bell. As shown in the map of the 50-node case in Fig.~\ref{fig:toronto_map}, Bell has better coverage in Western Toronto while weak in the Eastern. Rogers has a relatively balanced edge node geographical distribution. Then, we assume a virtual EIP owns all the edge nodes of Rogers and Bell (labeled as Combined EIP in Fig.~\ref{fig:extending}). Moreover, for fairness, we set all three EIPs to have the same total amount of resources (i.e., same storage and computation capacity). Fig.~\ref{fig:extending} presents the performance of different EIPs, and it can be seen that the combined EIP outperforms Rogers and Bell with 13.3\% and 10.6\% performance improvement, respectively. The black curve further illustrates this result with a cloud resource utilization ratio. The cloud utilization ratio of the Combined EIP is the highest, which indicates that the optimal provisioning schedule could be more efficient in the edge nodes extending scenario, as more cloud resources are utilized and the overall cost is reduced.

\subsubsection*{\textbf{Adaptive Vertical Resource Allocation}}To test the effectiveness of the dynamic service provisioning algorithm in edge federation, we calculate the resource utilization ratio of the services with seven different latency requirements (i.e., the latency requirement groups in Table.~\ref{table_requirement}). The results are shown in Fig.~\ref{fig:utilization}. We can see that when the requirement becomes more and more strict, the edge resource utilization ratio of all the services is increasing. This indicates that when facing the varying latency requirements, the algorithm truly realizes the dynamic resource utilization adjustment between edge and cloud resources, i.e., utilizing more edge resources under the strict requirement. 

The above all experimental results show that edge federation indeed solves the difficulties and challenges presented in Sec.II. It performs particularly effective under the heavy load and strict latency requirements, which fully match the needs of the latency-critical and resource-intensive smart services and show the value of our model in the real network environment.

\section{Discussion and Future Works}\label{sec:discussion}

\subsection{Determining the Length of the Time Slot} The performance of edge federation could be affected by the length of the scheduling time slot. Compared with the fixed contract model in 30-node case, we present a preliminary result of edge federation with different lengths of the time slot (e.g., 30mins, 60mins, and 120mins.). As shown in Fig.~\ref{fig:diff_timeslot}, we can observe that: a shorter time slot (i.e., the higher rescheduling frequency) can yield better performance. For instance, the 30mins, 60mins, and 120mins can achieve 20.5\%, 19.5\%, and 18.2\% cost savings, respectively. For simplicity, we assume that the overhead of each rescheduling $V_{re}$ is the same, which includes the costs of computation and communication. Hence, compared with the setting of 120mins, the settings of 30mins and 60mins obtain 2.3\% and 1.3\% cost saving gain, while incurring $4\times$ ($120/30$) and $2\times$ ($60/30$) times overhead loss, respectively. This may not be a cost-efficient deal. In this work, we assume that only the additional rescheduling cost is less than the saving improvement of the total cost, EIPs may be willing to take a higher rescheduling frequency. Thus, the rescheduling frequency largely depends on the rescheduling cost $V_{re}$ and the total service cost $V$. 

Automatically determining the length of the time slot is an interesting and promising idea. One possible solution could be applying reinforcement learning to the scheduling, which aims to make better decisions by learning from experiences through trial and error interactions with the environment~\cite{peng2018deepmimic}. In our context, based on the historical data, the reinforcement learning agent can adapt quickly to the dynamic service demands change, and thus be readily applied to unique network environments and architectures. Then, the rescheduling frequency can be flexibly adjusted according to the actual network environments and service requirements. We leave this as an open problem for our future works.

\subsection{Determining the Optimal Controlling Scale}
Rather than solving problems in the specific scenario, the edge federation is a general resource management model for the macro edge-computing scenario. The edge federation is operated in a centralized control manner, which could enable the most cost-efficiency service management for EIPs and provide the guaranteed QoS and QoE for the ESP and the end user, respectively. 

One of the critical issue for the centralized management is the scale of the controlling area, which greatly determined by the factors in geography (e.g., different time zones may affect the prediction accuracy, different users in different areas may have different behavior patterns.), business environment (e.g., unique business policies in different countries and regions.), etc.. According to these factors, the centralized control in a city, a country or a strongly related region (e.g., the area of European union countries) can be more effective and robust. Traffic behaviors of these areas are more predictable and amenable to provide a mathematically well-grounded sizing solution.

\subsection{Designing the Suitable Algorithm}

The networking environment in this paper is quite complicated. We formulate the optimization problem in the edge federation by mainly considering 1) resource factors (e.g., the heterogeneous resources of communication, storage, and computation); 2) geo factors (e.g., distributed edge nodes and users); 3) traffic factors (e.g., heterogeneous services, varying service demands, different latency requirements). Then, what we should do is finding the optimal analytical solution by solving the optimization problem. The primary purpose of this paper is to prove that edge federation is more cost-efficiency than the existing solution. Additionally, from other perspectives, one can also design new algorithms or exploit the advantages of the optimization techniques to solve problems (e.g., latency minimization, etc.) in edge federation. We leave this point as an open issue.

\section{Related work}\label{related work}
The related work can be roughly divided into two categories, including the service placement method and the service provisioning model.

Service placement is a popular topic in mobile edge computing (MEC), which involves the content caching and the computation offloading. The content caching has been studied extensively to avoid frequent replication and enable faster access by placing a large volume of content based on the popularity~\cite{dan2014dynamic}. The caching of the multimedia content is a representative field of the content caching area. Many efforts have been made on the collaborative multi-bitrate video caching and processing in MEC network~\cite{tran2017collaborative},~\cite{li2018qoe}. To enhance the QoS and QoE, some works also seek the aid of network technologies (e.g., software-defined network~\cite{li2018virtual}, network function vituralization~\cite{borjigin2018broker}) to efficiently manage network caching resources and delivery service content. Recently, the emerging concept of \emph{In-Network Caching} has been proposed~\cite{llorca2013dynamic}. The basic idea of \emph{In-Network Caching} is that according to the contents' popularity, servers tend to cache some content passing through them and enable a global caching architecture. Based on such the strategy, each server may send the required content directly to the end users with a small round trip time.

The computation offloading mainly focuses on designing dedicated strategies for offloading partial even the entire task from an end device to edge servers. The major factors influence the offloading strategies including the characteristics of end devices and edge servers, such as the location~\cite{wang2014mobility}, energy~\cite{barbera2013offload}, and different optimization objectives (e.g., minimizing the cost~\cite{ma2017cost} or delay~\cite{liu2016delay}). Liu et al. propose a searching algorithm to find the optimal task scheduling policy to achieve the minimum average delay~\cite{liu2016delay}. Mao et al. develop a LODCO algorithm to minimize the execution delay and addressed the task failure as the performance metric~\cite{mao2016dynamic}. There is also some literature jointly consider the caching and offloading for maximizing the revenue of mobile network operator~\cite{zhou2017resource}. Different from these works mentioned above, our work considers a general multi-EIP scenario.

Although the service provisioning is a crucial issue for edge computing, there still lack sufficient studies. The most involved literature focuses on the integration between cloud and edge. Tong et al. design a hierarchical edge cloud architecture to alleviate the peak workload from end users~\cite{tong2016hierarchical}, and Xu et al. also propose a similar hierarchical architecture with in-memory caching function to enable an energy-efficient caching scheme~\cite{xu2018saving}. To minimize the cost of resources, Ma et al. propose a cloud-assisted framework in MEC, named \emph{CAME} \cite{ma2017cost}, by combing the queueing network and convex optimization theories. Villari et al. also propose a similar architecture call \emph{Osmotic Computing}, which aims to decompose the applications into microservices and enhance seamless cooperation between cloud and edge resources~\cite{villari2016osmotic}. It is true that literature~\cite{wang2017joint} considers the cooperation between cells in a cellular network and~\cite{chen2017exploiting} even study the D2D collaboration among edge devices. However, there still lacks much literature study the cooperation among edge servers, as pointed out in this paper.

Such a dilemma has already attracted considerable attention from industries, several organizations (e.g., OpenFog\footnote{https://www.openfogconsortium.org/}, EdgeComputingConsortium\footnote{http://en.ecconsortium.org/}) have been formed trying to find the effective network architecture and service provisioning model. To our best knowledge, this is the first step to consider the service provisioning model from the entire edge-computing environment. Our edge federation considers the service-provisioning problem among multiple EIPs and cloud with hard latency constraints.

\section{Conclusion}\label{conclusion}

In this paper, we proposed an integrated service provisioning model, named \emph{edge federation}, which considered a two-dimension integration between multiple EIPs, including the vertical and the horizontal. Over the edge federation, we formulated the provisioning process as an LP problem and took a variable dimension shrinking method to solve the large-scale optimization problem. Furthermore, for varying service demands, we proposed a dynamic service provisioning algorithm, \emph{SEE}, which dynamically updates schedules to enable an efficient service deployment. Via the trace-driven experiments conducted on the real-world base station map of Toronto, we demonstrated that our edge federation model can help EIPs save the overall cost by 23.3\%to 24.5\%, and 15.5\% to 16.3\%, compared with the existing fixed contract model and multihoming model, respectively.

\section{Acknowledgment}\label{ACK}
This work was supported in part by the National Natural Science Foundation of China under Grant U19B2024, Grant 61802421 and Grant 71571186, in part by National Key Research and Development Program under Grant 2018YFE0207600, and in part by the National Science Foundation of Hunan Province under Grant 2019JJ30029.

\bibliographystyle{IEEEtran}
\balance
\bibliography{references}

\end{document}